\documentclass[article]{aa}  

\usepackage{graphicx}
\usepackage[none]{hyphenat}
\usepackage[multiple]{footmisc}
\usepackage{xcolor}
\usepackage{subcaption}
\usepackage[colorlinks=true,allcolors=blue]{hyperref}
\bibpunct{(}{)}{;}{a}{}{,} 

\usepackage{txfonts}
\usepackage{gensymb}
\newcommand{\Ha}{$\mathrm{H}\alpha$\xspace}
\newcommand{\Hb}{$\mathrm{H}\beta$\xspace}
\newcommand{\Hg}{$\mathrm{H}\gamma$\xspace}
\newcommand{\Hd}{$\mathrm{H}\delta$\xspace}
\newcommand{\He}{$\mathrm{H}\epsilon$\xspace}
\newcommand{\OII}{$[\mathrm{O}~\textsc{ii}]$\xspace}
\newcommand{\OIIlines}{$[\mathrm{O}~\textsc{ii}]\,\lambda\lambda 3726,29$\xspace}
\newcommand{\OIIa}{$[\mathrm{O}~\textsc{ii}]\,\lambda 3726$\xspace}
\newcommand{\OIIb}{$[\mathrm{O}~\textsc{ii}]\,\lambda 3729$\xspace}
\newcommand{\NeIIIline}{$[\mathrm{Ne}\textsc{iii}]\,\lambda 3869$\xspace}

\newcommand{\OIII}{$[\mathrm{O}~\textsc{iii}]$\xspace}
\newcommand{\OIIIlines}{$[\mathrm{O}~\textsc{iii}]\,\lambda\lambda 4959,5007$\xspace}
\newcommand{\HeIline}{$\mathrm{He}~\textsc{i}\,\lambda 5876$\xspace}
\newcommand{\OIIIb}{$[\mathrm{O}~\textsc{iii}]\,\lambda 5007$\xspace}
\newcommand{\OIline}{$[\mathrm{O}~\textsc{i}]\,\lambda 6300$\xspace}
\newcommand{\NII}{$[\mathrm{N}~\textsc{ii}]$\xspace}
\newcommand{\NIIlines}{$[\mathrm{N}~\textsc{ii}]\,\lambda\lambda 6548,6584$\xspace}

\newcommand{\NIIb}{$[\mathrm{N}~\textsc{ii}]\,\lambda 6584$\xspace}
\newcommand{\SII}{$[\mathrm{S}~\textsc{ii}]$\xspace}
\newcommand{\SIIlines}{$[\mathrm{S}~\textsc{ii}]\,\lambda\lambda 6716,31$\xspace}
\newcommand{\SIIa}{$[\mathrm{S}~\textsc{ii}]\,\lambda 6716$\xspace}
\newcommand{\SIIb}{$[\mathrm{S}~\textsc{ii}]\,\lambda 6731$\xspace}

\begin{document} 

   \title{GA-NIFS: Co-evolution within a highly star-forming galaxy group at z$\sim$3.7 witnessed by JWST/NIRSpec IFS}

 \titlerunning{JWST/NIRSpec view of GS4891}
   \author{B. Rodr\'iguez Del Pino
          \inst{\ref{iCAB}}\thanks{e-mail: brodriguez@cab.inta-csic.es}
          \and
          M. Perna\inst{\ref{iCAB}}
          \and
          S. Arribas\inst{\ref{iCAB}}
          \and
          F. D'Eugenio\inst{\ref{iKav},\ref{iCav}}
          \and    
          I. Lamperti\inst{\ref{iCAB}}
          \and
          P. G.~P\'erez-Gonz\'alez\inst{\ref{iCAB}}
          \and
          H. \"Ubler\inst{\ref{iKav},\ref{iCav}}
          \and 
          A. Bunker\inst{\ref{iOxf}}
          \and 
          S. Carniani\inst{\ref{iNorm}}
          \and 
          S. Charlot\inst{\ref{iSor}}
          \and 
          R. Maiolino\inst{\ref{iKav},\ref{iCav}, \ref{iUCL}}          
          \and
          C. J. Willott\inst{\ref{iNRC}}
          \and 
          T. B{\"o}ker\inst{\ref{iESOba}}
          \and
          J. Chevallard\inst{\ref{iOxf}}
          \and 
          G. Cresci\inst{\ref{iOAA}}
          \and 
          M. Curti\inst{\ref{iESOge}}
          \and 
          G. C. Jones\inst{\ref{iOxf}}
          \and
          E. Parlanti\inst{\ref{iNorm}}
          \and 
          J. Scholtz\inst{\ref{iKav},\ref{iCav}}
          \and 
          G. Venturi\inst{\ref{iNorm}}
          }

   \institute{Centro de Astrobiolog\'ia, (CAB, CSIC--INTA), Departamento de Astrof\'\i sica, Cra. de Ajalvir Km.~4, 28850 -- Torrej\'on de Ardoz, Madrid, Spain\label{iCAB}
        \and 
        Kavli Institute for Cosmology, University of Cambridge, Madingley Road, Cambridge, CB3 0HA, UK\label{iKav}
        \and
        Cavendish Laboratory - Astrophysics Group, University of Cambridge, 19 JJ Thomson Avenue, Cambridge, CB3 0HE, UK\label{iCav}
        \and
        Department of Physics, University of Oxford, Denys Wilkinson Building, Keble Road, Oxford OX1 3RH, UK\label{iOxf}   
        \and
        Scuola Normale Superiore, Piazza dei Cavalieri 7, I-56126 Pisa, Italy\label{iNorm}
        \and
        Sorbonne Universit\'e, CNRS, UMR 7095, Institut d’Astrophysique de Paris, 98 bis bd Arago, 75014 Paris, France\label{iSor} 
        \and
        Department of Physics and Astronomy, University College London, Gower Street, London WC1E 6BT, UK\label{iUCL}
        \and 
        National Research Council of Canada, Herzberg Astronomy \& Astrophysics Research Centre, 5071 West Saanich Road, Victoria, BC V9E 2E7, Canada\label{iNRC}
        \and
        European Space Agency, c/o STScI, 3700 San Martin Drive, Baltimore, MD 21218, USA\label{iESOba}
        \and 
        INAF - Osservatorio Astrofisco di Arcetri, largo E. Fermi 5, 50127 Firenze, Italy\label{iOAA}
        \and 
        European Southern Observatory, Karl-Schwarzschild-Straße 2, 85748, Garching, Germany\label{iESOge}
   }

   \date{Received ... ; accepted ...}

  \abstract{We present NIRSpec IFS observations of a galaxy group around the massive GS\_4891 galaxy at $z\sim3.7$ in GOODS-South that includes two other two systems, GS\_4891\_n to the north and GS\_28356 to the east. These observations, obtained as part of the GTO Galaxy Assembly - NIRSpec IFS (GA-NIFS) program, allow us to study for the first time the spatially resolved properties of the interstellar medium (ISM) and the ionised gas kinematics of a galaxy at this redshift. Leveraging the wide wavelength range spanned with the high-dispersion grating (with resolving power R=2700) observations, covering from \OIIlines to \SIIlines, we explore the spatial distribution of the star formation rate, nebular attenuation, and gas metallicity, together with the mechanisms responsible for the excitation of the ionised gas. GS\_4891 presents a clear gradient of gas metallicity (as traced by 12 + log(O/H)) by more than 0.2~dex from the southeast (where a star-forming clump is identified) to the northwest. The gas metallicity in the less massive northern system, GS\_4891\_n, is also higher by 0.2~dex than at the centre of GS\_4891, suggesting that inflows of lower-metallicity gas might be favoured in higher-mass systems. The kinematic analysis shows that GS\_4891 presents velocity gradients in the ionised gas consistent with rotation. The region between GS\_4891 and GS\_4891\_n does not present high gas turbulence, which, together with the difference in gas metallicities, suggests that these two systems might be in a pre-merger stage. Finally, GS\_4891 hosts an ionised outflow that extends out to {$r_{\rm out}$=$1.5$~kpc} from the nucleus and reaches maximum velocities, {\bf $v_{\rm out}$}, of approximately 400~km/s. Despite entraining an outflowing mass rate of $\dot{M}_{\rm out}\sim4$~M$_{\odot}$/yr, the low associated mass-loading factor, $\eta\sim0.04$, implies that the outflow does not have a significant impact on the star formation activity of the galaxy. 
}


   \maketitle
%

\section{Introduction}
According to observations, the star formation rate density (SFRD) of the Universe peaked at around $3.5-6$~Gyr after the Big Bang ($1<z<2$) and about $25\%$ of the present-day stellar mass was already formed {{by then}} \citep{madauCosmicStarFormationHistory2014}. Consequently, galaxies observed at earlier epochs ($<3.5$~Gyr) are expected to be actively star-forming and to be undergoing processes 
such as galaxy mergers and/or the accretion of large amounts of gas and stars, which could lead to significant mass growth and a global increase in the SFRD. The evolutionary paths that these early galaxies follow can be constrained through the study of the imprints that different drivers of galaxy evolution leave on galaxies. In particular, the ongoing star formation activity, properties of the interstellar medium (ISM) such as gas metallicity, dust content, and ionisation conditions, as well as the global kinematics of galaxies, can be probed through the rest-frame optical emission. However, until recently, detailed, spatially resolved studies of galaxies at high redshifts have only been possible with ground-based near-infrared integral field spectroscopic observations accessing the \Ha line up to $z\sim2.6$ \citep{forsterschreiberSINSSurveySINFONI2009, stottKMOSRedshiftOne2016, forsterschreiberKMOS3DSurveyDemographics2019b, wisnioskiKMOS3DSurveyData2019, curtiKLEVERSurveySpatially2020} and the \OIIIb line up to $z\sim3.5$ \citep{troncosoMetallicityEvolutionMetallicity2014, turnerKMOSDeepSurvey2017b}. These previous works have revealed that galaxies at these epochs present a variety of metallicity gradients \citep{cresciGasAccretionOrigin2010,troncosoMetallicityEvolutionMetallicity2014,wuytsEvolutionMetallicityMetallicity2016, curtiKLEVERSurveySpatially2020}, with a high incidence of ionised outflows in massive galaxies {\citep{genzelSinsSurveyGalaxy2011, forsterschreiberKMOS3DSurveyDemographics2019b, guptaMOSELSurveyExtremely2023, llerenaIonizedGasKinematics2023}} but inconspicuous outflows in lower-mass ones \citep[e.g.][]{concasBeingKLEVERCosmic2022}, and an increase in the intrinsic velocity dispersion (higher pressure support) with redshift \citep{turnerKMOSDeepSurvey2017b, ublerEvolutionOriginIonized2019}. 
Unfortunately, comparable studies of galaxies beyond $z\sim3$ are impossible because the bright \Ha line is redshifted outside the near-infrared window observable from the ground.

Thankfully, after a successful launch and commissioning, the James Webb Space Telescope (JWST) \citep[JWST; ][]{gardnerJamesWebbSpace2023} has opened a new era in the exploration of the early epochs of galaxy evolution through its access to near- and mid-infrared wavelengths. Since the start of scientific operations, JWST has provided numerous results on the properties of galaxies beyond $z\sim3$, which, compared to lower-redshift ones, are found to be more metal-poor \citep{schaererFirstLookJWST2022a, curtiJADESInsightsLowmass2023, nakajimaJWSTCensusMassMetallicity2023, trumpPhysicalConditionsEmissionline2023}, display harder ionising spectra \citep{cameronJADESProbingInterstellar2023, sandersExcitationIonizationProperties2023}, and exhibit higher electron densities \citep{isobeRedshiftEvolutionElectron2023, reddyJWSTNIRSpecExploration2023}.

Despite the wealth of information already obtained from spectroscopic observations of large samples of galaxies, most studies have been based on their integrated emission \citep[e.g.][]{cameronJADESProbingInterstellar2023, carnianiJADESIncidenceRate2023,sandersExcitationIonizationProperties2023}, and a spatially resolved analysis of their properties is still missing. In this regard, the advent of integral field spectroscopy (IFS) instrumentation aboard JWST provides, for the first time, the possibility of spatially resolving the full rest-frame optical suite of emission lines, from \OII to \Ha+ \NII+ \SII, in galaxies beyond $z>3$, to study the physical properties of galaxies at early epochs. In particular, the NIRSpec instrument \citep{jakobsenNearInfraredSpectrographNIRSpec2022a} provides IFS capabilities \citep{bokerNearInfraredSpectrographNIRSpec2022} to study, in the near-infrared regime, the rest-frame optical emission from galaxies up to $z\sim9$. However, until now, only a few spatially resolved studies have been performed on galaxies within this redshift range \citep[e.g.][]{pernaUltradenseInteractingEnvironment2023a, ublerGANIFSMassiveBlack2023a, wylezalekFirstResultsJWST2022}. 

In this paper, we shed more light on the internal properties of high-redshift galaxies through the spatially resolved study, using NIRSpec IFS observations, of a galaxy group at $z\sim3.7$ that includes the massive, star-forming galaxy GS$\_$4891. Previously observed in CANDELS \citep{groginCANDELSCosmicAssembly2011, koekemoerCANDELSCosmicAssembly2011} as part of the GOODS-South campaign, GS$\_$4891 lies at a favourable redshift to cover a wide range of spectral features from \OIIlines to \SIIlines in a single NIRSpec spectroscopic band at a high spectral resolution. 

\begin{figure}
\centering
\includegraphics[width=0.48\textwidth]
{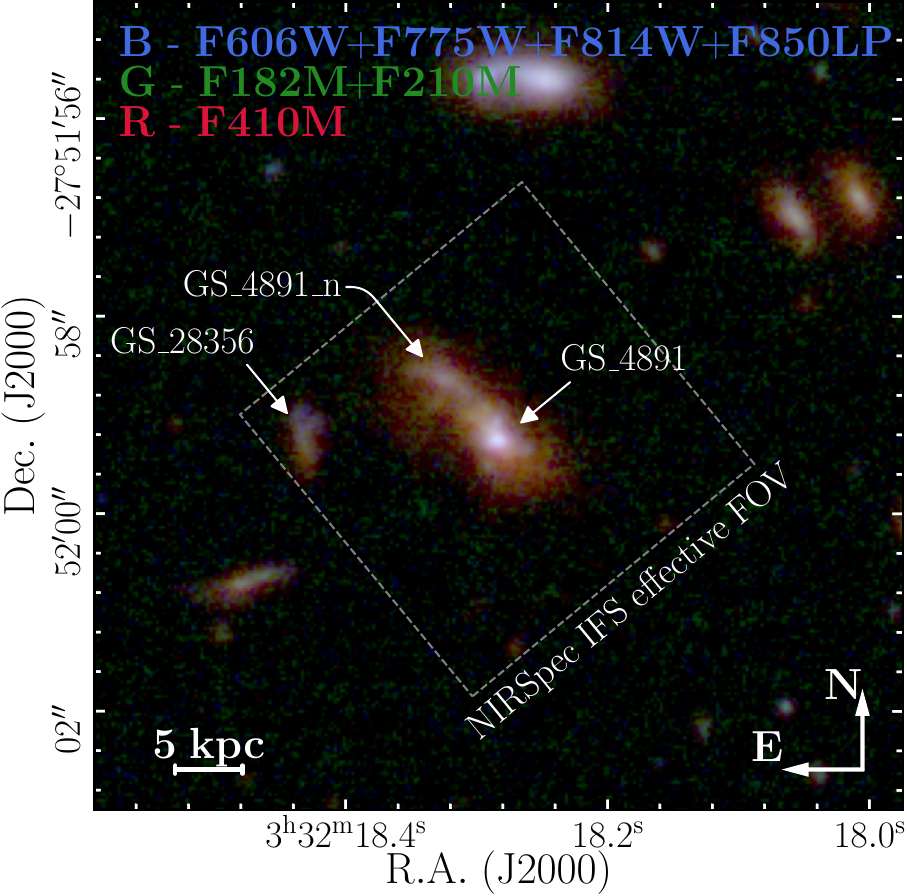}
\caption{False-colour image of the field around GS\_4891 highlighting the position of the NIRSpec IFS field of view. Blue colours are a combination of ACS/HST images \citep{giavaliscoGreatObservatoriesOrigins2004}, whereas green and red correspond to NIRCam images \citep{williamsJEMSDeepMediumband2023}.}
\label{fig:imaging}
\end{figure}

This paper is organised as follows. In Section~\ref{sec:data_reduction} we describe the NIRSpec observations, the reduction, and the analysis of the data. In Section~\ref{sec:results} we present the results obtained on the ISM properties and kinematics in the GS\_4891 system, together with a discussion of their implications. Finally, the summary of the main results and concluding remarks are presented in Section~\ref{sec:conclusions}.

Throughout this paper, we adopt the initial mass function (IMF; $0.1-100$ M$_{\odot}$) of \citet{chabrierGalacticStellarSubstellar2003} and a flat $\Lambda$CDM cosmology with $H_{\rm 0}$~=~70~km~s$^{\rm -1}$~Mpc$^{\rm -1}$, $\Omega_{\rm \Lambda}$~=~0.7, and $\Omega_{\rm m}$~=~0.3. Emission lines are referred to using their rest-frame air wavelengths, although for the analysis we use their vacuum wavelengths. 

\begin{figure*}
\centering
\includegraphics[width=0.65\textwidth]
{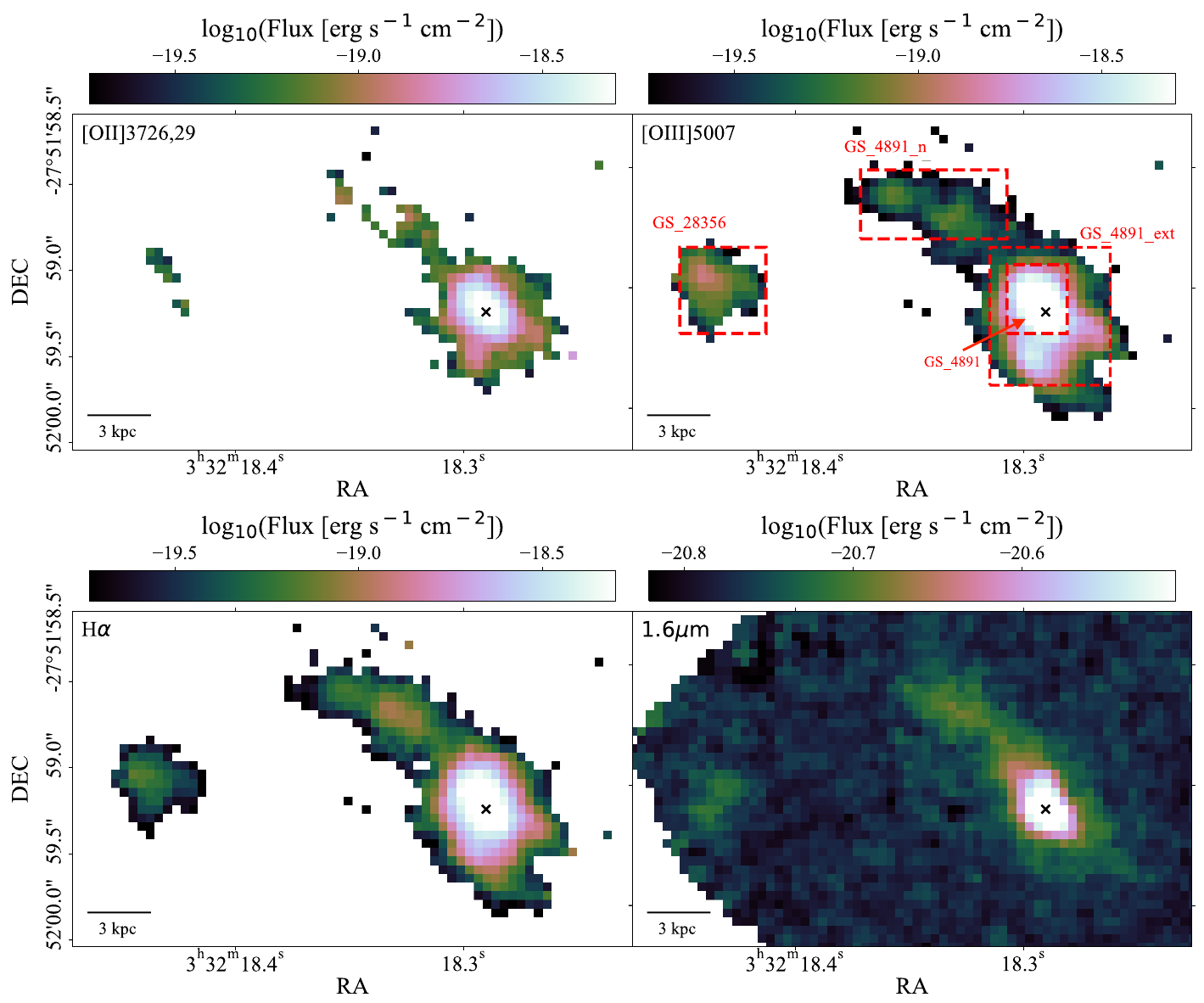}
\caption{Maps of integrated fluxes in the emission lines \OIIlines, \OIIIb, and \Ha, {and the continuum around 1.6$\mu$m} for the GS\_4891 system. The three main galaxies identified in the system are highlighted with red rectangles in the \emph{top-right} panel. The black cross indicates the centroid of continuum emission (see text for details). North is up and east is to the left. }
\label{fig:flux_maps}
\end{figure*}

\begin{figure}
\begin{center}
\includegraphics[width=0.49\textwidth]{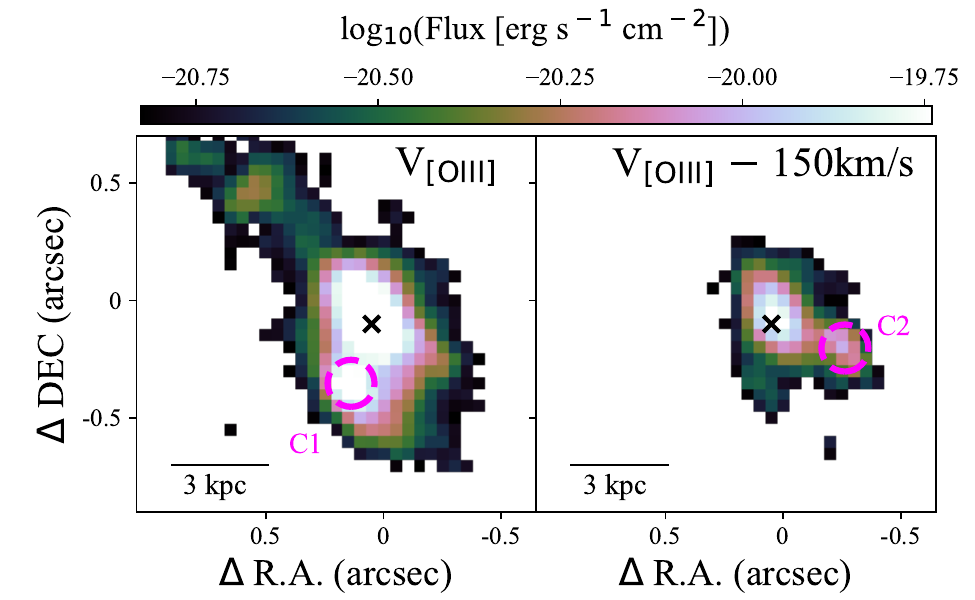}
\caption{Flux maps in windows of two spectral pixels ($\sim$~50~km/s wide) centred at $25$~km/s and $-170$~km/s with respect to the position of \OIIIb at the systemic redshift ($z$=3.7027) where the C1 and C2 separate clumps (magenta circles) are identified. The origin of the coordinates corresponds to R.A. 3$^{\rm h}$32$^{\rm m}$18.29$^{\rm s}$, Dec. $-27$\degree51$'$59.17$''$, J2000. North is up and east is to the left.}
\label{fig:clumps_flux_maps}
\end{center}
\end{figure}

\section{Observations and data reduction}
\label{sec:data_reduction}

\subsection{JWST/NIRSpec IFS observations}
\label{subsec:observations}

The data presented in this paper are part of the NIRSpec IFS GTO program ‘Galaxy Assembly with NIRSpec IFS’ (GA-NIFS; PIs: S. Arribas, R. Maiolino) and are included in proposal \#1216 (PI: N. Luetzgendorf). Observations of GS$\_$4891 (R.A. 3$^{\rm h}$32$^{\rm m}$18.29$^{\rm s}$, Dec. $-27$\degree51$'$59.17$''$, J2000) were executed on August 13 2022, with the IFS mode of NIRSpec covering a contiguous area of 3.1\arcsec~$\times~$3.2\arcsec with a native spaxel size of 0.1\arcsec \citep{bokerNearInfraredSpectrographNIRSpec2022, rigbySciencePerformanceJWST2023}. IFS observations of GS\_4891 were obtained at high- (R2700) and low-resolution (R100) configurations, applying a medium (0.5\arcsec) cycling pattern of eight dithers. The R2700 observations were obtained with a total integration time of 15872.7 seconds (4.4 hours) using the grating-filter pair G235H/F170LP, which provides a spectral resolution, $R\sim1900-3500$ between $1.7~\mu$m and $3.15~\mu$m \citep{jakobsenNearInfraredSpectrographNIRSpec2022a}. A total of 3968.2 seconds (1.1  hours) were devoted to the R100 observations ($R\sim30-330$ between $0.6~\mu$m and $5.3~\mu$m).  

Since GS\_4891 resides in the GOODS-South field, it has been previously observed with multi-wavelength observations, including WFC3/HST and Spitzer/IRAC. Based on these previous observations, GS\_4891 has a spectroscopic redshift of 3.7027, a stellar mass of approximately $1.3\times10^{10}$~${\rm M}_{\odot}$, and a star formation rate (SFR) of 49.5~M$_{\odot}$/yr \citep{guoCANDELSMultiwavelengthCatalogs2013, perez-gonzalezSpitzerViewEvolution2005}. 

In Figure~\ref{fig:imaging} we show a false-colour image of the group of galaxies highlighting the field of view of the NIRSpec IFS observations presented here. The main galaxy, GS$\_$4891, has a very close companion to the north, which we refer to as GS$\_$4891$\_$n, and another galaxy to the east at a projected distance of $\sim15$~kpc  identified in CANDELS as GS$\_28356$ (photo-$z$=3.646). The green and red channels correspond to NIRCam  F182M+F210M and F410M images, respectively, from the JWST Extragalactic Medium-band Survey \citep[JEMS;][]{williamsJEMSDeepMediumband2023}, whereas the blue channel is a combination of ACS/HST archival data \citep[F606W, F775W, F814W, F850LP;][]{giavaliscoGreatObservatoriesOrigins2004}, re-processed following \citet{illingworthHubbleLegacyFields2016} and \citet{whitakerHubbleLegacyField2019}. 

\subsection{Data reduction}
\label{subsec:data_reduction}

The raw data were reduced with the JWST calibration pipeline version 1.8.2 under CRDS context jwst\_1068.pmap. Several modifications to the default reduction were introduced to improve the data quality; these are described in detail in \citet{pernaUltradenseInteractingEnvironment2023a}. The final cube was combined using the ‘drizzle’ method, for which we used an official patch to correct a known bug.\footnote{https://github.com/spacetelescope/jwst/pull/7306.} The main analysis in this paper is based on the combined R2700 cube with a pixel scale of 0.05$\arcsec$. During the analysis we noticed that the noise provided in the datacube (`ERR' extension) is underestimated compared to the actual noise in the data. For this reason, the noise vector in each spaxel was re-scaled based on the standard deviation in the continuum in regions free of emission lines, increasing the noise by a factor that varies with wavelength and ranges between 1.5 and 2. No background subtraction was performed on the R2700 data because in this paper we focus on the analysis of emission lines and any background contribution is taken into account in the spectral modelling. The same steps were applied to reduce the R100 data, with the addition of the background subtraction using spaxels away from the sources. 

\begin{figure*}
\centering
\includegraphics[width=\textwidth]
{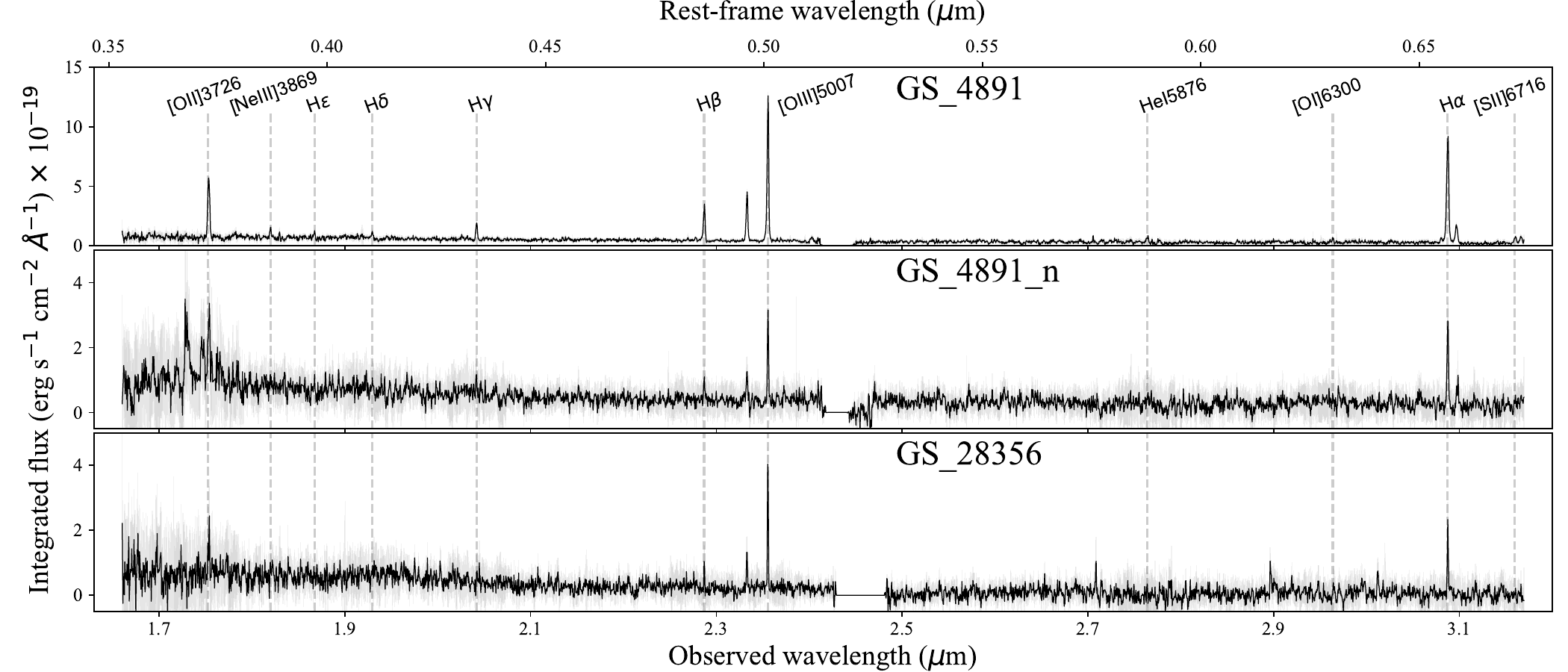}
\caption{Integrated spectra of GS$\_$4891 and the two other systems, GS$\_$4891$\_$n and GS$\_$28356, identified in the NIRSpec/IFS observations. The integrated spectra were obtained by adding all the spaxels within the apertures highlighted in Figure~\ref{fig:flux_maps}.}
\label{fig:integrated_spectra}
\end{figure*}

\subsection{Analysis}
\label{subsec:analysis}

\subsubsection{R2700 data - Emission line analysis}
The main emission lines (\OIIlines, \Hb, \OIIIb, \Ha, \NIIlines, and \SIIlines) were modelled, in the spectrum of each spaxel and in the integrated spectrum of the regions defined in Section~\ref{subsec:first_look}, with a set of individual Gaussian functions tied together to have the same kinematics. In this fit, the ratios \OII$\lambda3729$/$\lambda3726$ and \SII$\lambda6716$/$\lambda6731$ are constrained within [0.3839,1.4558] and [0.4375, 1.4484], respectively, corresponding to the theoretical limits of low- (1~cm$^{-3}$) and high- (10$^5$~cm$^{-3}$) density regimes estimated by \citet{sandersMOSDEFSurveyElectron2016}. We also set a minimum value for the \Ha/\Hb flux ratio of 2.86,
which corresponds to an electron temperature, T$_{\rm e}$~=~$10^4$~K, and an electron density, N$_{\rm e}$~=~$100$~cm$^{-3}$, for a case B recombination \citep{osterbrockAstrophysicsGaseousNebulae2006}. 

We explored the presence of additional kinematic components in the spectra by including a second set of Gaussian functions, with a larger velocity dispersion, in a separate fit to the brighter \OIIIb and \Ha lines, also tying their kinematics together. The two-component model was
preferred when the Bayesian information criterion statistics \citep{schwarzEstimatingDimensionModel1978} was significantly better (> 60) than for the
one-component model and the broad component presented a
maximum flux that was at least two times larger than the scatter
in the neighbouring continuum. In these cases, the rest of the emission lines (\OIIlines, \Hb and \NIIlines) were modelled again, forcing them to have the same two kinematic components as in the fit to \OIIIb and \Ha, and only allowing the flux to vary in each component. The narrow kinematic component of the fit was associated with the galaxy and used to study its ISM properties and gas kinematics; the broad component was used to explore the presence and properties of a galactic outflow (see Sect.~\ref{subsubsec:results_outflow}). 

The signal-to-noise (S/N) in each line was computed as $TotalFlux$ / $RMS\times \sqrt{2\times FWHM}$, where $TotalFlux$ is the summed flux within $2\times FWHM$ (full width at half maximum) of the line and the RMS is measured in windows of 20 pixels in the continuum around the emission lines. {The adopted $FWHM$ was taken from an initial fit performed on the brightest lines (i.e. \OIIIb and \Ha) on a spaxel-by-spaxel basis}. In order to extract physical information from regions with low S/N, in some of the emission lines the spectra from different spaxels were combined following the Voronoi binning method from \citet{cappellariAdaptiveSpatialBinning2003}. It should be noted that the binning changes depending on the target emission line; in particular, analyses that require the \NIIb line (e.g. N2 metallicity indicator, excitation mechanisms) demand larger bins than those that use \Hb (e.g. nebular extinction, R2 and R3 metallicity indicators). Errors in the physical parameters derived in this paper were computed with Markov Chain Monte Carlo techniques using the \textsc{EMCEE} software by \citet{foreman-mackeyEmceeMCMCHammer2013b}, assuming uniform priors. 
The reported 1-$\sigma$ uncertainties on each quantity were calculated as half the difference between the 16th and 84th percentiles of the posterior probability density distribution. The spectral modelling took into account the wavelength-dependent spectral resolution of NIRSpec \citep{jakobsenNearInfraredSpectrographNIRSpec2022a}.


\subsubsection{R100 data - Stellar mass estimation}
Previous stellar mass estimates of GS\_4891 were based on aperture photometry and also included the galaxy to the north, GS\_4891\_n (see Figure~\ref{fig:imaging}), since they appeared to be two connected structures in HST imaging. In this work we study them separately as we find evidence that they might be different structures in an early stage of interaction. Thus, we needed to estimate their individual stellar masses. 

The SED fitting to estimate stellar masses was done using the R100 NIRSpec IFS data (0.6$\mu$m to 5.3$\mu$m). For GS\_4891, the modelling was performed at the spaxel level in the same region used to model the kinematics of the system (see central panel in the middle row of Figure~\ref{fig:barolo} in Appendix~\ref{a:3dbarolo}), whereas for GS\_4891\_n we used the spaxels within the region demarcated in Figure~\ref{fig:flux_maps}. The method is described in \citet{perez-gonzalezCEERSKeyPaper2023} and \citet{deugenioFastrotatorPoststarburstGalaxy2023}. Briefly, the clear-prism NIRSpec/IFS observations were compared to stellar population models from the \citet{bruzualStellarPopulationSynthesis2003} library, assuming a star formation history described by a delayed exponential characterised by a timescale, $\tau$ (taking values from 1~Myr to 1~Gyr in 0.1~dex steps), and age, $t_0$ (ranging from 1~Myr to the age of the Universe at the redshift of the galaxy). The stellar metallicity, $Z$, was left as a free parameter, enabling us to take all the discrete values provided by the   \citet{bruzualStellarPopulationSynthesis2003} library from 2\% to 2.5 times solar. Nebular (continuum and line) emission was taken into account, asin \citet{perez-gonzalezStellarPopulationsLocal2003,perez-gonzalezStellarMassAssembly2008}. 
The attenuation of the stellar and nebular emission was modelled with a \citet{calzettiDustContentOpacity2000}
law, with $A_{\rm V}$ values ranging from 0 to 3 magnitudes in 0.1~mag steps. The stellar mass was obtained by scaling the mass-normalised stellar model to the spectrum.

{Integrating the light in the regions around GS\_4891\_ext and GS\_4891\_n shown in the top-right panel of Figure~\ref{fig:flux_maps}, we obtain stellar masses of $7.1\pm0.3\times10^9$~${\rm M}_{\odot}$ and $1.3$~$\pm$~$0.1$~$\times$~$10^9$~${\rm M}_{\odot}$, respectively. We note that the sum of these stellar masses is slightly lower than the previous estimations that included the two systems ($1.3\times10^{10}$~${\rm M}_{\odot}$); however, we ascribe this disagreement to small differences in the area encompassed by our individual apertures and the one used in the HST data}.

\begin{table*}
\caption{{Measured line fluxes and derived integrated physical properties from the main emission lines identified in the integrated spectra of GS\_4891, the internal ionised gas clumps, C1 and C2, and its total extended emission, GS\_4891\-ext, and the neighbouring galaxies GS\_4891\_n and GS\_28356  (Figure~\ref{fig:integrated_spectra})}. Fluxes were not corrected for extinction.}
\begin{center}
\begin{tabular}{lcccccc}
\hline
\hline
\rule{0pt}{2.5ex}  
&   \multicolumn{6}{c}{Line Fluxes ($\times$ 10$^{\rm -20}$ erg s$^{-1}$ cm$^{-2}$)} \\
\hline    
\rule{0pt}{2.5ex}  
& GS\_4891 & GS\_4891\_C1 & GS\_4891\_C2 & GS\_4891\_ext & GS\_4891\_n & GS\_28356\\
\hline
\rule{0pt}{2.5ex}  
\OIIa & $648.8\pm81.7$ & $66.9\pm11.6$ & $48.4\pm10.9$ & $1270.3\pm214.3$ & $244.3\pm99.7$ & $144.1\pm89.9$ \\
\OIIb & $680.6\pm180.8$ & $55.6\pm21.8$ & $53.4\pm19.2$ & $1239.6\pm440.6$ & $232.3\pm127.0$ & $129.7\pm94.8$ \\
\NeIIIline & $156.9\pm32.1$ & $29.3\pm5.4$ & $-$ & $-$ & $232.3\pm127.0$ & - \\
\Hd & $103.8\pm34.7$ & $25.2\pm10.8$ & $-$ & $-$ & $232.3\pm127.0$ & - \\
\Hg & $266.3\pm25.1$ & $29.4\pm9.3$ & $-$ & $482.0\pm103.3$ & $232.3\pm127.0$ & - \\
\Hb & $654.8\pm25.1$ & $61.8\pm5.4$ & $46.0\pm4.8$ & $1304.0\pm211.2$ & $108.9\pm36.6$ & $83.2\pm21.6$ \\
\OIIIb & $2467.5\pm35.5$ & $309.2\pm8.6$ & $142.4\pm9.4$ & $4555.6\pm102.3$ & $410.8\pm56.1$ & $449.1\pm30.1$ \\
\HeIline & $133.9\pm41.5$ & $-$ & $-$ & $-$ & $410.8\pm56.1$ & - \\
\Ha & $2485.5\pm33.4$ & $183.1\pm9.5$ & $131.6\pm10.9$ & $4913.1\pm299.5$ & $521.1\pm52.7$ & $253.4\pm37.1$ \\
\NIIb & $408.0\pm21.1$ & $14.1\pm6.0$ & $-$ & $681.6\pm223.8$ & $108.8\pm43.7$ & - \\
\SIIa & $162.6\pm55.8$ & $-$ & $-$ & $380.9\pm296.0$ & $-$ & $-$ \\
\SIIb & $162.0\pm64.7$ & $-$ & $-$ & $376.2\pm313.1$ & $-$ & $-$ \\
\hline
\hline
\end{tabular}
\rule{0pt}{2.5ex}  
\centerline{Integrated physical properties}
\begin{tabular}{lcccccc}
\hline
\rule{0pt}{2.5ex}  
& GS\_4891 & GS\_4891\_C1 & GS\_4891\_C2 & GS\_4891\_ext & GS\_4891\_n & GS\_28356\\
\hline
\rule{0pt}{2.5ex}  
$z$ & 3.7031 & 3.7026 & 3.7010 & 3.7029 & 3.7033 & 3.7032\\
$\sigma$~(km s$^{-1}$) & $100.7\pm1.2$ & $59.8\pm1.7$ & $59.1\pm4.1$ & $100.9\pm2.3$ & $65.1\pm8.3$ & $30.7\pm6.5$\\
A$_{\rm V}$ (mag) & $1.1\pm0.0$ & $0.1\pm0.0$ & $0.0\pm0.0$ & $1.1\pm0.0$ & $2.0\pm0.1$ & $0.2\pm0.1$\\
SFR (M$_{\odot}$yr$^{-1}$) & $37.5\pm0.6$ & $1.4\pm0.1$ & $0.9\pm0.1$ & $72.7\pm4.9$ & $14.4\pm1.6$ & $2.2\pm0.3$ \\
$\Sigma_{\rm SFR}$ (M$_{\odot}$yr$^{-1}$kpc$^{-2}$) & $6.6\pm0.1$ & $0.9\pm0.1$ & $0.6\pm0.1$ & $1.6\pm0.1$ & $1.0\pm0.1$ & $0.2\pm0.0$\\
12+log(O/H) & $8.4_{-0.1}^{+0.1}$ & $8.2_{-0.3}^{+0.2}$ & $8.3_{-0.5}^{+0.2}$ & $8.4_{-0.2}^{+0.1}$ & $8.5_{-0.1}^{+0.1}$ & $8.2_{-0.5}^{+0.3}$\\
logU &  $-2.5_{+0.0}^{-0.2}$ & $-2.5_{+0.0}^{-0.3}$ & $-2.5_{+0.0}^{-0.4}$ & $-2.5_{+0.0}^{-0.4}$ & $-3.1_{+0.1}^{-0.7}$ & $-2.8_{0.2}^{-0.9}$\\
N$_{\rm e}$ (cm$^{-3}$)    & $776\pm307$    & -    & -   & & -   & \\

\hline
\hline
\end{tabular}
\end{center}
\label{table:integrated_properties}
\end{table*}

\section{Results and discussion}
\label{sec:results}

\subsection{The complex neighbourhood around GS$\_$4891}
\label{subsec:first_look}

{In Figure~\ref{fig:flux_maps} we show the maps of the distribution of the total \OIIlines, \OIIIb, and \Ha fluxes, together with the continuum flux in the NIRSpec IFS R100 data in a wavelength window equivalent to the HST F160W/WFC3 filter, tracing the rest-frame UV emission. The black cross in this figure marks the centroid of continuum emission computed with a 2D Gaussian fit}. The three galaxies display spatially resolved emission in the two lines over several kiloparsecs. The red rectangles shown in the top-right panel of Figure~\ref{fig:flux_maps} were chosen to encompass most of the emission from each system and to study their individual properties. The integrated NIRSpec G235H/F170LP spectra from these apertures are shown in Figure~\ref{fig:integrated_spectra}, which demonstrates that the emission lines are robustly detected with high S/N in a wide spectral range. The similar observed wavelength of the emission lines in their spectra confirms that they reside at the same redshift and are part of a galaxy group. The three galaxies exhibit clear detections of strong emission lines such as the \OIIlines doublet, \OIIIlines, and the hydrogen Balmer recombination lines, \Hb and \Ha. Additionally, GS$\_$4891 also displays other emission lines such as higher-order Balmer transitions lines (\Hg, \Hd, \He) as well as \NeIIIline, \HeIline, \OIline, \NIIlines, and \SIIlines.

Apart from the three main structures, the map of the \OIIIb flux displays evidence of substructure to the southeast of GS$\_$4891, which is less clearly seen in \Ha. The presence of this and further substructure is investigated in more detail through the flux maps shown in Figure~\ref{fig:clumps_flux_maps} that correspond to windows of two spectral pixels ($\sim$~50~km/s wide) centred at $25$~km/s (left panel) and $-170$~km/s (right panel) with respect to the position of \OIIIb at the systemic redshift ($z$=3.7027). The plot on the left shows a clump of ionised gas (C1) to the southeast of GS\_4891 at the same velocity, whereas an additional clump (C2) is seen in the southwest moving at approximately $-150$~km/s with respect to the main galaxy. These two clumps can also be identified in the composite HST+NIRCam image shown in 
Figure~\ref{fig:imaging} through their blue colours, indicative of strong rest-frame UV radiation likely coming from young stars. This composite image also shows a compact blue clump in the north of the galaxy, closer to its centre than the other two clumps, which does not stand out in the NIRSpec data. The observed internal substructure suggests either in situ star formation or ongoing minor mergers.

\subsection{Properties of the ISM}
\label{subsec:ISM_properties}

\subsubsection{Electron densities}
\label{subsubsec:results_ne}

\begin{figure}
\begin{center}
\includegraphics[width=0.45\textwidth]{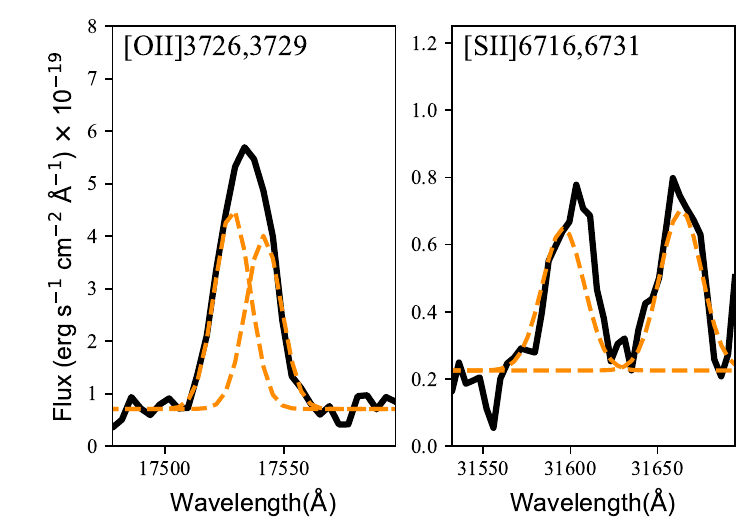}
\caption{Simultaneous spectral modelling (orange lines) of the observed \OIIlines and \SIIlines (black lines) to constrain the electron density of the ionised gas in the integrated spectrum of GS$\_$4891.}
\label{fig:electron_density}
\end{center}
\end{figure}

The detection of the \OIIlines and \SIIlines doublets in the integrated spectra of GS\_4891 (see Figure~\ref{fig:integrated_spectra}) offers the possibility of computing the electron density of the ionised gas \citep{osterbrockAstrophysicsGaseousNebulae2006}. Although each individual set of emission lines can be used for this computation, the detection of both of them can be leveraged to provide a better constraint on the electron density. Assuming a typical electron temperature for HII regions, T$_{\rm e}$~=~$10^4$~K, and following the expressions derived by \citet{sandersMOSDEFSurveyElectron2016},
 the \OIIlines and \SIIlines in the integrated spectra of GS\_4891 were simultaneously modelled, forcing them to have the same kinematics and to agree on the value of the electron density.  In this fit, we also constrained the line ratios within the same theoretical limits from \citet{sandersMOSDEFSurveyElectron2016} stated in Section~\ref{subsec:analysis}. This modelling, shown in Figure~\ref{fig:electron_density}, yields an electron density, $n_{\rm e}$ = $776\pm307$~cm$^{-3}$, whereas the independent modelling of \OIIlines and \SIIlines gives electron densities of $533\pm357$~cm$^{-3}$ and $357\pm309$~cm$^{-3}$, respectively. The three estimates are consistent within the (large) uncertainties and provide a range of electron densities compatible with our data. 

 Our high $n_{\rm e}$ measurements in GS\_4891 are in agreement with an overall increase in electron density with redshift observed in galaxies at $z<3$ \citep{steidelStrongNebularLine2014a,sandersMOSDEFSurveyElectron2016, kashinoFMOSCOSMOSSURVEYSTARFORMING2017,daviesKMOS3DSurveyInvestigating2021a}. However, our values are also higher than the median electron densities, $n_{\rm e}$~$\sim$~$300$~cm$^{-3}$, measured using \OIIlines in a sample of ten galaxies with masses around 10$^9$~M$_{\odot}$ at $4\lesssim$~$z$~$\lesssim6$ by \citet{isobeRedshiftEvolutionElectron2023} and much higher than in the sample of 24 massive (mean log(M$_{\star}$/M$_{\odot}$)=9.86) galaxies at $2.7<z<6.3$ studied in \citet{reddyJWSTNIRSpecExploration2023}, where they find a mean $n_{\rm e}<$100~cm$^{-3}$ using \SIIlines. Nevertheless, our estimates of the electron density are still within the large scatter [$40-600$~cm$^{-3}$] observed in these previous works.

In their work on a total of 48 galaxies at $2.7<z<6.3$,
\citet{reddyJWSTNIRSpecExploration2023} find that the electron density increases for higher values of the O32 (\OIIIb/(\OIIlines)), with the highest densities, $n_{\rm e}> 500$~cm$^{-3}$, being associated with O32~$>$~2. Such results are interpreted as a consequence of the positive correlation between the SFR surface density ($\Sigma_{\rm SFR}$, whose increment leads to higher electron densities) and the ionisation parameter (log$U$, traced by O32), both found to increase with redshift. At high masses they find that galaxies have low electron densities, $n_{\rm e}< 100$~cm$^{-3}$, and O32 $<6$, due to the lower $\Sigma_{\rm SFR}$ of more massive systems. In the case of GS\_4891, the extinction-corrected estimate of O32, 1.36 is lower than the O32 values [$2-8$] measured for electron densities $> 500$~cm$^{-3}$ in \citet{reddyJWSTNIRSpecExploration2023}, whereas $\Sigma_{\rm SFR}$~=~6.6$\pm0.1$~M$_{\odot}$~yr$^{-1}$~kpc$^{-2}$, estimated within the aperture highlighted in Figure~\ref{fig:flux_maps}, is in agreement with the range they obtain [$2-20$~M$_{\odot}$~yr$^{-1}$~kpc$^{-2}$] in the high-density regime. 

Overall, a comparison with previous works indicates that GS\_4891 presents an electron density that is higher than expected for a massive galaxy at $z$=$3.7$, although the uncertainties are large. A possible explanation could be the presence of an ionised outflow in the central regions of GS\_4891 (see Sect.~\ref{subsubsec:results_outflow}), since galactic outflows are generally found to have higher-density gas 
\citep{arribasIonizedGasOutflows2014,villarmartinTriggeringMechanismProperties2014,mingozziMAGNUMSurveyDifferent2019a,rodriguezdelpinoPropertiesIonizedOutflows2019}.
The electron density measurements in GS\_4891 provide an estimate of the physical conditions of the ISM, reinforcing the assumption of a lower limit of \Ha/\Hb=2.86 (a value that changes to 2.85 only at N$_{\rm e}$~=~$10^4$~cm$^{-3}$, \citealt{grovesBalmerDecrementSloan2012}) that is used in the spectral modelling of our data (Section~\ref{subsec:analysis}) and the computation of the nebular extinction (Section~\ref{subsubsec:extinction}).

\subsubsection{Extinction}
\label{subsubsec:extinction}
\begin{figure}
\begin{center}
\includegraphics[width=0.45\textwidth]{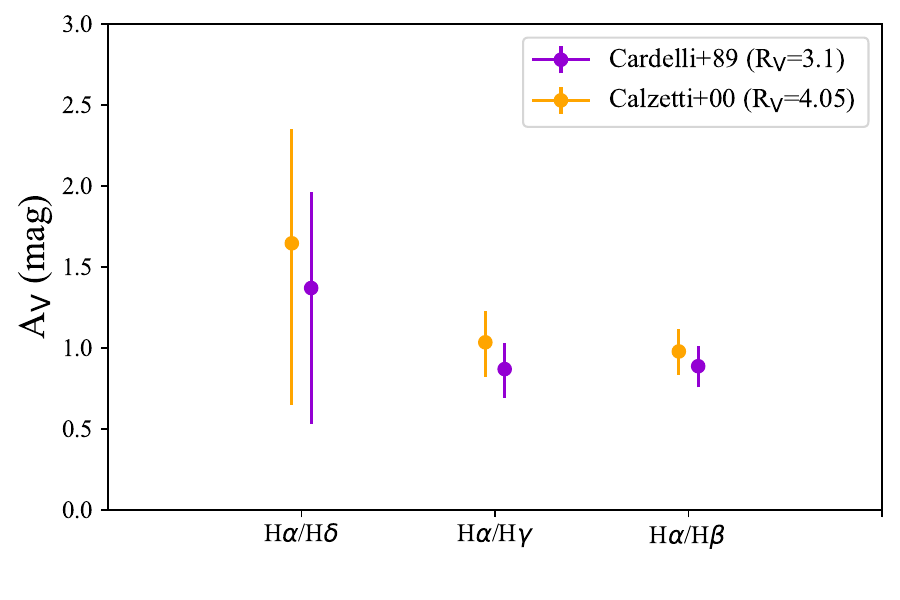}
\caption{A$_{\rm V}$ values for the integrated spectrum of GS$\_$4891 obtained using different pairs of hydrogen Balmer lines. The purple and orange points were obtained using the attenuation curves from \citet{cardelliRelationshipInfraredOptical1989} and \citet{{calzettiDustContentOpacity2000}}, respectively.} 
\label{fig:balmer_ratios}
\end{center}
\end{figure}

The availability of several hydrogen Balmer emission lines in the integrated spectrum of GS\_4891 enables us to evaluate the consistency of their flux ratios with standard attenuation curves. Using the \Hd, \Hg, \Hb and \Ha fluxes we derive the nebular extinction, $A_{\rm V}$, assuming the attenuation laws from \citet[][$R_{\rm V}$=3.1]{cardelliRelationshipInfraredOptical1989} and \citet[][$R_{\rm V}$=4.05]{calzettiDustContentOpacity2000}. The $A_{\rm V}$ values derived from the flux ratios between \Ha and the higher-order Balmer lines for each attenuation law are shown in Figure~\ref{fig:balmer_ratios}. All estimations of $A_{\rm V}$ present good agreement within the errors and are consistent with $A_{\rm V}$~$\sim$~$0.9-1$~mag, indicating that in GS\_4891 both attenuation laws are compatible. Such consistency with local attenuation laws was also observed in stacks of star-forming galaxies at $1.4<z<2.6$ by \citet{reddyMOSDEFSurveyFirst2020a} and, together with our results, suggests that at high redshifts (at least up to $z\sim3.7$) the effects of dust along the line of sight are similar to those in the local Universe. Thus, throughout this work we computed the nebular extinction from the Balmer decrement (\Ha / \Hb) following the method described in \citet{dominguezDustExtinctionBalmer2013}, which assumes Calzetti's attenuation law and a minimum \Ha/\Hb~=~2.86. 

\begin{figure*}
    \centering
    \includegraphics[width=.70\linewidth]{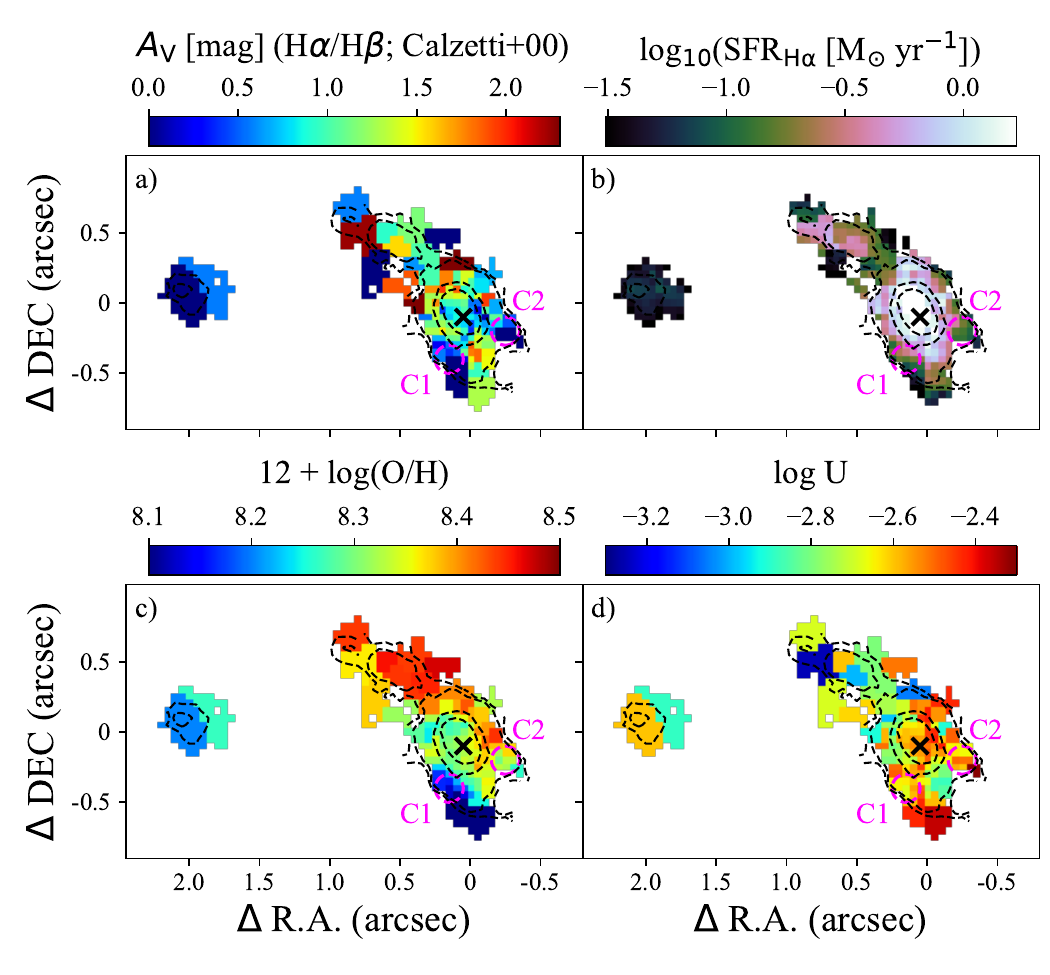}
\caption{Maps of the main ISM parameters derived for the GS\_4891 galaxy group from emission line analysis. \emph{a)} Nebular extinction, $A_{\rm V}$, derived from the Balmer decrement (\Ha/\Hb). \emph{b)} SFRs (M$_{\odot}$yr$^{-1}$) estimated from the \Ha luminosity. \emph{c)} Gas oxygen abundance obtained with the R2, R3, R23, and O32 metallicity calibrations from \citet{curtiNewFullyEmpirical2017}. \emph{d)} Ionisation parameter, $U$, computed with the gas metallicity and the \OIIlines lines following Equation~5 in \citet{diazChemicalAbundancesIonizing2000}. The data were spatially binned to have a $S/N\geq3$ in the emission lines involved. The dashed black contours correspond to the \Ha emission, whereas the black cross indicates the centroid of continuum emission.}
\label{fig:maps_derived_parameters}
\end{figure*}

A map of the nebular extinction for the galaxies in our system is shown in panel a) of Figure~\ref{fig:maps_derived_parameters}, after binning the data to have $S/N\geq3$ in \Hb. The integrated $A_{\rm V}$ of each system within the apertures defined in Figure~\ref{fig:flux_maps} are included in Table~\ref{table:integrated_properties}. Throughout the whole system, the nebular extinction varies by more than two magnitudes, including regions with almost no extinction. GS\_4891 presents $A_{\rm V}$ of roughly one magnitude at the centre, consistent with its integrated value. In the case of GS\_4891\_n, the integrated nebular extinction is quite high ($A_{\rm V}$~$\sim$~2~mag), but displays a large internal scatter. Interestingly, the region between GS\_4891 and GS\_4891\_n presents high extinction, $A_{\rm V}\geq1.5$, despite it not hosting significant star formation activity (black contours corresponding to the \Ha emission). On the contrary, clumps C1 and C2 present very low extinction, A$_{\rm V}<0.5$. The spatially resolved and integrated A$_{\rm V}$ estimates were used to correct for nebular attenuation the measured line fluxes in the combined spectra. {However, for the map of the SFR shown in Figure~\ref{fig:maps_derived_parameters}, we kept a spaxel-by-spaxel approach, applying the same extinction correction to all the spaxels within a given bin.} 

\begin{figure*}
\begin{center}
\includegraphics[width=0.95\textwidth]{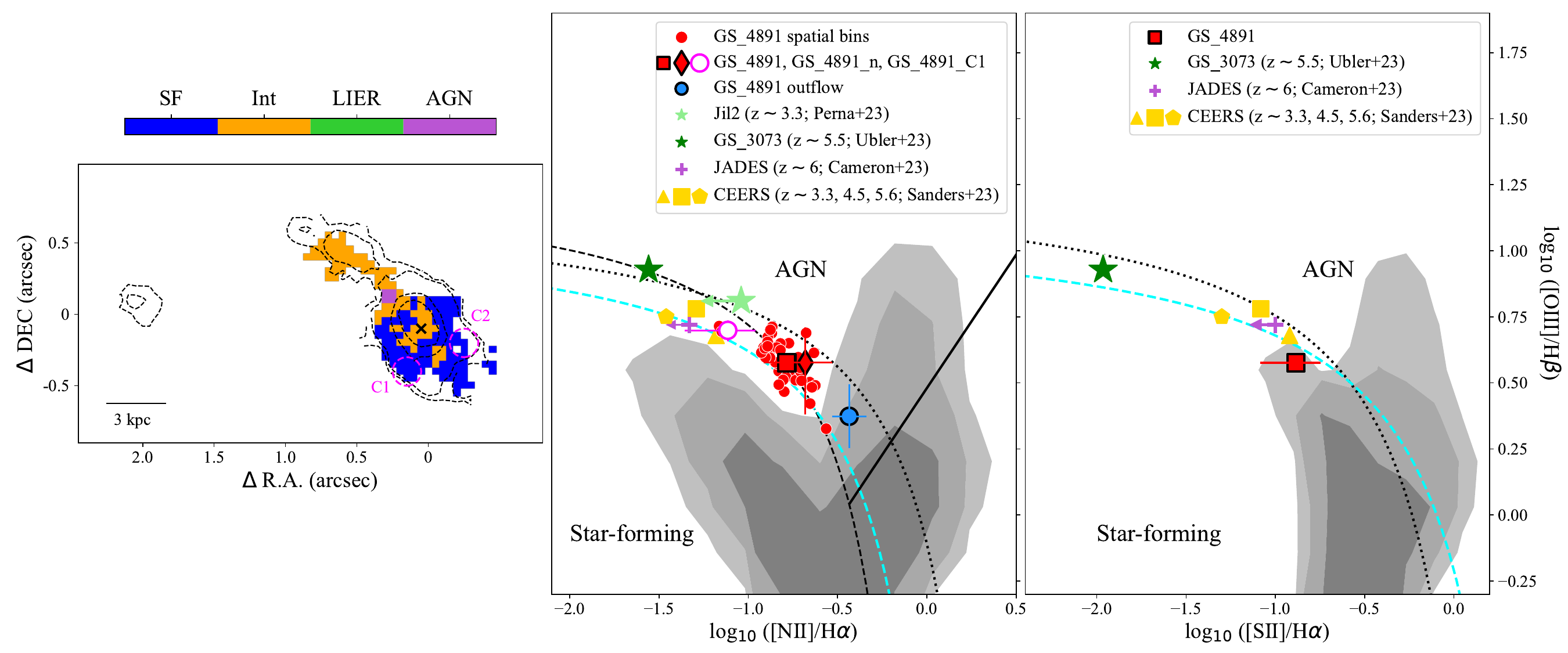}
\caption{\emph{Left}: Map of excitation mechanisms throughout GS$\_$4891 and GS$\_$4891$\_$n according to the location in the \NII-based BPT diagnostic diagram. The dashed black contours correspond to the \Ha emission, whereas the black cross indicates the centroid of continuum emission. \emph{Centre} and \emph{right}: \NII- and \SII-based BPT diagrams containing the spatially binned regions (red dots), the integrated regions of the system where the relevant lines are detected with $SN\geq3$ (red square and diamond, and pink circle), and the broad kinematic component associated with the ionised outflow identified in GS\_4891 (blue circle). {Clump C2 is not shown in any of the diagrams due to the low S/N in \Hb and \NII}. As a reference, we have added the stack of galaxies at $z\sim6$ from \citet[][purple plus]{cameronJADESProbingInterstellar2023}, the stacks in different high-redshift bins from \citet[][yellow symbols]{sandersExcitationIonizationProperties2023}, the obscured AGN at $z\sim3.3$, `Jil2', from \citet[][lightgreen star]{pernaUltradenseInteractingEnvironment2023a}, and the broad-line AGN, GS\_3073, at $z\sim5.5$ studied in \citet[][darkgreen star]{ublerGANIFSMassiveBlack2023a}. The shaded grey contours demarcate the regions in the diagram encompassed by 70, 80, and 90 percent of the local SDSS sample \citep{abazajianSeventhDataRelease2009}. Dashed \citep{kauffmannHostGalaxiesActive2003b} and dotted black lines \citep{kewleyTheoreticalModelingStarburst2001} demarcate the separation between ionisation coming from star formation and AGN activity based on studies of local galaxies (see Section~\ref{subsubsec:results_BPT} for details). The cyan lines correspond to the best-fit curves to the star-forming population at $z\sim2.3$ from \citet{stromNebularEmissionLine2017}.}
\label{fig:BPT}
\end{center}
\end{figure*}

\subsubsection{Excitation mechanisms}
\label{subsubsec:results_BPT}

The NIRSpec IFS observations of GS\_4891 allow us, for the first time, to spatially resolve the sources of ionisation in a galaxy at $z\sim3.7$ using rest-frame optical emission lines. This is an intermediate redshift between the well-studied range $z<2.5$ via ground-based observations \citep[e.g.][]{stromNebularEmissionLine2017} and the recently analysed cases at $z>5$ with JWST \citep{cameronJADESProbingInterstellar2023, sandersExcitationIonizationProperties2023}. Here, we explore the sources of ionisation throughout the galaxies in the GS\_4891 system by using the standard diagnostic diagram \citep{baldwinClassificationParametersEmissionline1981a} that includes the rest-frame optical emission line ratios \OIIIb/\Hb and \NIIb/\Ha. We also consider an alternative diagram that uses \SIIlines instead of \NIIb \citep{veilleuxSpectralClassificationEmissionLine1987} for the integrated spectrum of GS\_4891 (\OIline could also be used instead but is barely detected with $S/N<2$). The \NII- and \SII-based Baldwin, Phillips, and Terlevich (BPT) diagrams are shown in Figure~\ref{fig:BPT} together with a map of the distribution of the different sources of ionisation in our system following the \NII-based classification. {We note that clump C2 is not shown in the BPT diagram because the \NIIb line has $S/N<3$}. The diagrams include reference lines from studies of local galaxies that demarcate the regimes that could be explained considering only ionisation from star formation \citep[below the black, dashed line from][]{kauffmannHostGalaxiesActive2003b} and the one where theoretical models require the presence of active galactic nucleus (AGN) activity to explain the ionisation state of the gas \citep[black, dotted lines from][]{kewleyTheoreticalModelingStarburst2001}. The `intermediate' regime (between both lines) corresponds to ionisation that cannot be explained either by star formation or AGN activity alone \citep{cidfernandesAlternativeDiagnosticDiagrams2010}. The solid black line is aimed at separating AGNs from LIERs \citep[low-ionisation emission-line regions;][]{monreal-iberoLINERlikeExtendedNebulae2006, belfioreSDSSIVMaNGA2016}, as was suggested by \citet{cidfernandesAlternativeDiagnosticDiagrams2010}. In addition, we also show as a cyan line the best-fit curves tracing the location of the star-forming population at $z\sim2.3$ from \citet{stromNebularEmissionLine2017}.

According to these demarcation lines, there is no evidence for the presence of an AGN in either GS\_4891 or GS\_4891\_n, since all the points (with the exception of one region away from the nucleus of GS\_4891 that is probably spurious) lie below the \citet{kewleyTheoreticalModelingStarburst2001} line. However, we have to take into account that standard BPT diagrams are designed for local galaxies and have been proven to be less reliable in constraining excitation mechanisms in higher-redshift galaxies due to their lower metallicities. In fact, photoionisation models \citep[e.g. ][]{nakajimaDiagnosticsPopIIIGalaxies2022} predict that low-metallicity AGNs have lower \OIII/\Hb and \NII/\Ha ratios, locating them in regions of the BPT diagram that correspond to `intermediate' and star-forming regimes. Such an effect can be observed in high-redshift AGNs selected through other tracers such as the detection of a broad-line region in the Balmer emission lines \citep{harikaneJWSTNIRSpecFirst2023, kocevskiHiddenLittleMonsters2023, maiolinoJADESDiversePopulation2023,oeschJWSTFRESCOSurvey2023} or alternative diagnostic diagrams involving other emission lines that require harder ionising radiation such as {$\mathrm{He}\textsc{ii}\,\lambda 4686$} \citep{ublerGANIFSMassiveBlack2023a}. As an example, we show in Figure~\ref{fig:BPT} (dark green star) the low-metallicity AGN (GS\_3073; $Z_{\rm gas}$/$Z_\odot\sim0.21$) at $z\sim5.5$ studied in \citet{ublerGANIFSMassiveBlack2023a} that, despite being classified as star-forming following the standard BPT diagram, was confirmed to be an AGN based on the presence of a broad-line region and the diagnostic diagram involving the {$\mathrm{He}\textsc{ii}\,\lambda 4686$} line. We also show the obscured AGN at $z\sim3.3$, `Jil2' (light green star), that lies very close to the demarcation line from \citet{kewleyTheoreticalModelingStarburst2001} but for which \citet{pernaUltradenseInteractingEnvironment2023a} inferred a {$\mathrm{He}\textsc{ii}\,\lambda 4686$} consistent with AGN ionisation. Although these previous works demonstrate that the BPT diagram must be used with caution in high-redshift studies, our NIRSpec IFS spectra do not present evidence of a broad-line region in either \Hb or \Ha, and the higher ionisation emission line {$\mathrm{He}\textsc{ii}\,\lambda 4686$} is not detected (Fig.~\ref{fig:integrated_spectra}). Moreover, the gas metallicities we measure are $Z_{\rm gas}$/$Z_\odot\sim0.35-0.6$ (Section~\ref{subsubsec:results_metallicities}), a regime where excitation diagnostics using the BPT diagram should still be reliable. Finally, the \SII-based BPT classification (right panel in Figure {\ref{fig:BPT}) for the integrated spectrum of GS\_4891 is also consistent with ionisation from star formation.

\begin{figure}
    \centering
    \includegraphics[width=.98\linewidth]{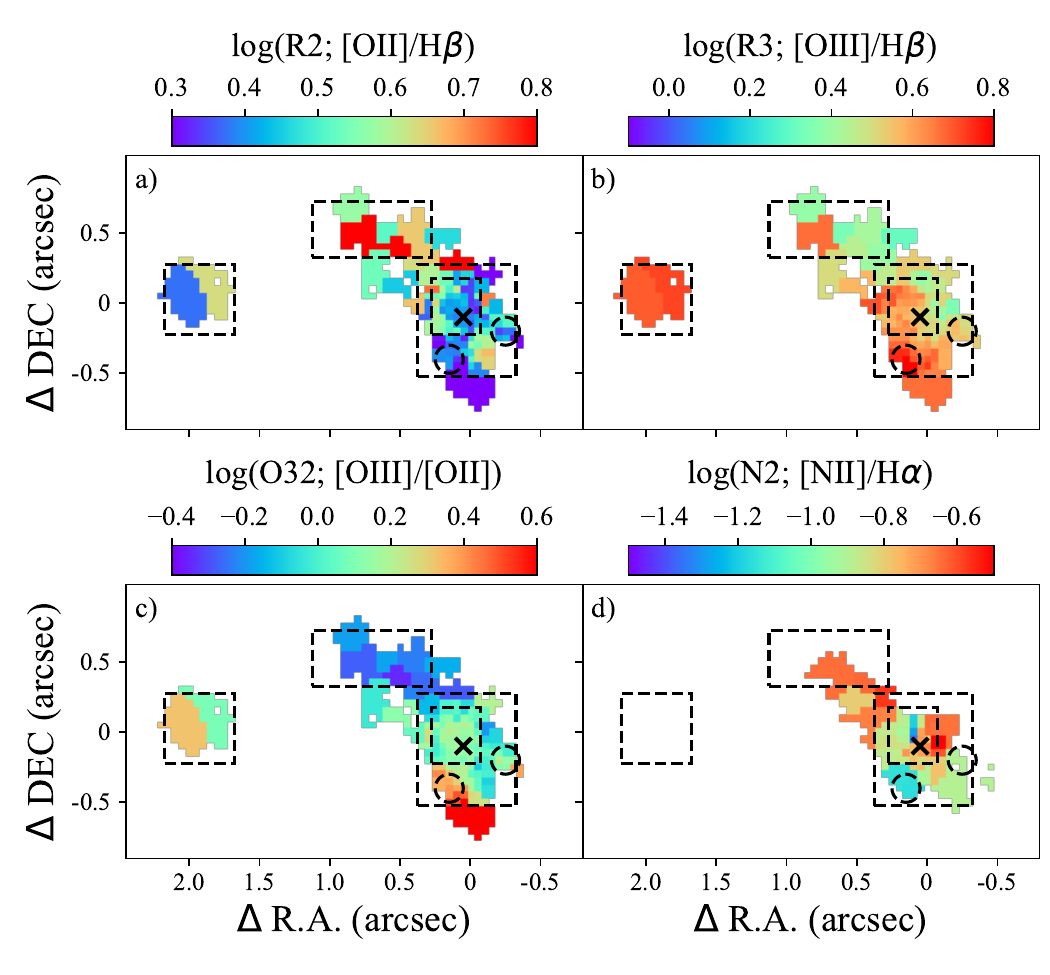}
\caption{Maps of emission line ratios that trace gas metallicity throughout the GS\_4891 system. The data have been spatially binned to have a $S/N\geq3$ in the emission lines involved. The black cross indicates the centroid of continuum emission, whereas dashed contours in the right panel correspond to the \Ha emission. Black boxes and circles highlight the individual systems identified in Section~\ref{subsec:first_look}.}
\label{fig:map_lineratios}
\end{figure}


Figure~\ref{fig:BPT} shows that the spatially resolved and integrated regions in our system are clearly offset from the parameter space covered by local star-forming galaxies \citep[SDSS;][]{abazajianSeventhDataRelease2009}.
If our system were at $z\sim0$, such deviations could be interpreted as signs of harder ionisation mechanisms such as those originating in shocks induced by galactic outflows \citep[e.g.][]{hoSAMIGalaxySurvey2014} or by mergers \citep[e.g.][]{pernaMUSEViewArp2202020}. In fact, as is described in Section~\ref{subsubsec:results_outflow}, we find evidence of an ionised outflow in the central regions of GS\_4891 oriented towards the northwest, a direction in which there is an increase in the global velocity dispersion (right panel of Figure~\ref{fig:kinematic_maps}) and in the \NIIb/\Ha ratio (panel d in Figure~\ref{fig:map_lineratios}), a correlation expected in regions ionised by shocks \citep[e.g.][]{hoSAMIGalaxySurvey2014}. In the case of GS\_4891\_n, the \NIIb/\Ha is even higher and it presents velocity dispersions ($\sigma\sim60$~km/s) that are lower but still consistent with the turbulence expected from shocks \citep{johnstonBPTNewMultiDimensional2023}. Moreover, the substructure observed in GS\_4891 and the close distance to GS\_4891\_n are compatible with ongoing or past mergers that could induce shocks. However, despite shocks possibly contributing to the observed line ratios, it is already well established that star-forming galaxies at high redshifts tend to display higher \OIIIb/\Hb at fixed \NII/\Ha ratios than local galaxies due to changes in their physical conditions \citep{brinchmannNewInsightsStellar2008, kewleyTHEORETICALEVOLUTIONOPTICAL2013, shapleyMOSDEFSURVEYEXCITATION2015, sandersMOSDEFSurveyElectron2016, steidelReconcilingStellarNebular2016, stromNebularEmissionLine2017}. This tendency is clearly illustrated by the location in Figure~\ref{fig:BPT} of the star-forming population at $z\sim2.3$ from \citet{stromNebularEmissionLine2017} and the higher-redshift stacks from \citet{cameronJADESProbingInterstellar2023} and \citet{sandersExcitationIonizationProperties2023}. The GS\_4891 data points cluster together slightly above the star-forming population at $z\sim2.3$, although they are still consistent within the intrinsic scatter (Figure 5 in \citealt{stromNebularEmissionLine2017}). Interestingly, clump 
C1 is offset from the rest of the GS\_4891 system due to its lower metallicity and higher ionisation, being the only region close to the $z\sim3.3$ stack (yellow triangle). However, we note that this data point from \citet{sandersExcitationIonizationProperties2023} corresponds to a wide redshift range, $2.7<z<4.0$, with a larger scatter in the BPT diagram (see Figure 3 in their paper) that is consistent with the line ratios observed in the GS\_4891 system. 

In summary, the GS\_4891 system occupies a region in the BPT diagram in between $z\sim2$ and $z>4$ galaxies, consistent with an evolution in the ionisation state with redshift. Moreover, the possible merging activity of GS\_4891 and the presence of an ionised outflow indicate that shocks could also contribute to the observed line ratios.

\subsubsection{Star formation rates}
\label{subsubsec:SFRs}

The results from the previous section demonstrate that the ionisation in the GS\_4891 is dominated by star formation. Therefore, the extinction-corrected \Ha fluxes can be used to estimate the ongoing SFRs following the calibrations listed in \citet{kennicuttStarFormationGalaxies1998}, consistent with a Salpeter IMF \citep{salpeterLuminosityFunctionStellar1955}. A map of the SFR is shown in panel b) of Figure~\ref{fig:maps_derived_parameters}, whereas the integrated values of each individual region are included in Table~\ref{table:integrated_properties}. As expected, the distribution of the SFR is very similar to that of the \Ha emission, peaking at the central regions of GS$\_$4891 and GS$\_$4891$\_$n, with a gap of low SFR between both systems. The total integrated SFR (\Ha) of these two systems is approximately 52~M$_{\odot}$~yr$^{-1}$, which is very similar to the one previously estimated from SED modelling of HST+IRAC/Spitzer photometry, 49.5~M$_{\odot}$~yr$^{-1}$ \citep{perez-gonzalezSpitzerViewEvolution2005}, also using a Salpeter IMF. However, as can be noted, using the individual apertures highlighted in Figure~\ref{fig:flux_maps} we are not including all the \Ha flux of the system. If we sum all the \Ha flux in the GS\_4891 and GS\_4891\_n structures, we obtain a total SFR of $\sim125$~M$_{\odot}$~yr$^{-1}$, doubling previous SED estimations. This difference in the SFRs estimated using \Ha and SED fitting might be due to the fact that \Ha emission traces star formation on shorter time scales than the UV and IR emission traced in the SED fitting, implying that GS$\_$4891 is experiencing a burst of star formation at the time of observations. Taking into account their stellar masses and SFRs, 
GS\_4891 and GS\_4891\_n would lie approximately 0.34~dex and 0.43~dex, respectively, above the main-sequence at $z=3.7$ \citep{speagleHighlyConsistentFramework2014}, indicating that 
these are starburst systems. Regarding GS\_28356, the \Ha-based SFR, $2.2\pm0.5$~M$_{\odot}$~yr$^{-1}$, is the same as the one previously estimated from SED fitting, 2.2~M$_{\odot}$~yr$^{-1}$.

\subsubsection{Gas metallicities}
\label{subsubsec:results_metallicities}
The study of the spatially resolved metal content of the ISM is critical to identifying regions in a galaxy that have been more metal-enriched due to subsequent episodes of star formation and those where diluted, pristine gas from the circumgalactic medium is being accreted. Leveraging the detection of several emission lines throughout our spatially resolved spectra, we can apply well-established calibrations \citep[e.g.][]{curtiNewFullyEmpirical2017} to constrain the oxygen abundance in the GS\_4891 group of galaxies. After correcting for dust attenuation, we use flux ratios between specific emission lines that correspond to standard metallicity indicators: R2 (\OIIlines/\Hb), R3 (\OIIIb/\Hb), O32 (\OIIIb / \OIIlines), R23 ((\OIIlines + \OIIIlines)/\Hb), and N2 (\NIIb / \Ha). The spatial distributions of four of these indicators, obtained after spatially binning the data to have $S/N\geq3$ in the emission lines involved (\Hb for the first three panels and \NIIb for the last one), are shown in Figure~\ref{fig:map_lineratios}. The metallicity calibrations from \citet{curtiNewFullyEmpirical2017} for the R2, R3, R23, and O32 parameters were fitted together to generate the oxygen abundance map presented in panel c) of Figure~\ref{fig:maps_derived_parameters} (see \citealt{curtiJADESInsightsLowmass2023} for details). The N2 indicator is not included in this estimation as it requires coarser spatial binning. The integrated oxygen abundances in each system were also estimated using the R2, R3, R23, and O32 indicators, with the addition of N2 for GS\_4891 and C1 where \NIIb is detected. Their values are presented in Table~\ref{table:integrated_properties}. We note that, since GS\_4891 is at $z=3.7$, the calibrations provided by \citet{sandersDirectT_ebasedMetallicities2023} based on galaxies at $2<z<9$ should, in principle, be more appropriate to computing the metallicities than the ones from \citet{curtiNewFullyEmpirical2017} that are based on local galaxies. However, the calibrations from \citet{sandersDirectT_ebasedMetallicities2023} do not provide good constraints at the regime of relatively high oxygen abundances (12+log(O/H) > 8.3) estimated in the GS\_4891 system. 


The three galaxies in the group display a wide range of oxygen abundances, from 12+log(O/H)~$\sim8.1$ to $\sim8.5$. In particular, GS\_4891 presents a clear gradient in metallicity from the south (including clump C1) to the northwest. The value at the centre, 12+log(O/H)~$\sim8.3$, is 0.2dex higher than in the south and $\sim0.15$dex lower than in the northwest. Regarding the drop in metallicity at the south, the low velocity dispersion ($\sim50$~km/s) estimated from the kinematic analysis performed in Sect.~\ref{subsubsec:results_gas_kinematics} (see also Figure~\ref{fig:kinematic_maps}) suggests that this is a region where lower-metallicity gas is likely being accreted through a minor merger. The positive gradient observed from the centre of GS\_4891 to the northwest resembles those previously reported in galaxies {at $z\geq3$ by \citet{curtiKLEVERSurveySpatially2020} and  \citet{cresciGasAccretionOrigin2010}, and recently at $z\sim7$ by \citet{arribasGANIFSCoreExtremely2023}}, which have generally been ascribed to the dilution of metallicity in the central regions as a consequence of new episodes of star formation fuelled by cold flows of pristine gas \citep{dekelFormationMassiveGalaxies2009}. Overall, the gradients observed in the oxygen abundance map are also visible in the individual line ratios presented in Figure~\ref{fig:map_lineratios}. In particular, the N2 indicator, which was not included in the estimation of the oxygen abundance and which correlates with metallicity  \citep{pettiniOIIINIIAbundance2004, curtiNewFullyEmpirical2017}, also displays a clear increase from the southeast to the northwest of GS\_4891. 

In addition to the observed metallicity gradient, probably the most striking result is the difference of approximately 0.2dex between the metallicities at the centre of GS\_4891 and GS\_4891\_n, given that the former is four times more massive and that it consequently should be more efficient at retaining metals \citep[e.g.,][]{tremontiOriginMassMetallicityRelation2004a}. In fact, while GS\_4891 has an integrated metallicity (12+log(O/H)=8.4) in agreement with the mass-metallicity relation at $z\sim3.3$ \citep{sandersMOSDEFSurveyEvolution2021a} given its mass ($5.5\pm0.6\times10^9$~${\rm M}_{\odot}$), the metallicity of GS\_4891\_n (12+log(O/H)=8.5) is $\geq0.3$dex higher than the mean value in galaxies of similar mass ($1.3$~$\pm0.1$~$\times$~$10^9$~${\rm M}_{\odot}$). A possible explanation for this result is that the most massive galaxy, GS\_4891, is the main beneficiary of inflows of lower-metallicity gas following the baryon cycle that maintains the mass-metallicity relation. Such a scenario is supported by the higher SFR density (the same as the SFR in Figure~\ref{fig:flux_maps}) in GS\_4891 that is expected in the case of the accretion of pristine gas \citep{ceverinoGasInflowMetallicity2016}}. Furthermore, the higher metallicity in GS\_4891\_n could be a consequence of gas stripping and `starvation' \citep[or `strangulation'][]{larsonEvolutionDiskGalaxies1980a, pengStrangulationPrimaryMechanism2015} by the surrounding environment. Considering GS\_4891\_n as a satellite galaxy and GS\_4891 as the central one, the difference in their metallicities resembles the findings at lower redshifts ($z<0.1$) where satellite galaxies are invariably more metal-rich than centrals of the same mass \citep[e.g.][]{pasqualiGasphaseMetallicityCentral2012a, pengDependenceGalaxyMassmetallicity2014, lianSDSSIVMaNGAEnvironmental2019, schaeferSDSSIVMaNGAEvidence2019}. Two scenarios have been proposed to explain this trend: either satellites accrete relatively more metal-rich gas (pre-processed by their more massive centrals, e.g.  \citealt{pengDependenceGalaxyMassmetallicity2014, guptaChemicalPreprocessingCluster2018a}), or, alternatively, satellites are more metal-rich due to a combination of starvation (which truncates gas accretion), ram-pressure stripping (which preferentially removes low-metallicity gas situated at large radii), and a lower impact of metal-loaded outflows \citep[e.g.][]{pasqualiGasphaseMetallicityCentral2012a, rodriguezdelpinoImpactEnvironmentalEffects2023}.
In the case of GS\_4891\_n, we see enrichment of a galaxy relative to its own central, which {could} favour the second scenario. The metallicity difference between GS\_4891 and its satellite could also explain the positive radial metallicity gradient observed from the centre to the northwest of GS\_4891: if the satellite already passed the pericentre, some of the high-metallicity gas to the northwest of GS\_4891 could be stripped of material originally belonging to GS\_4891\_n. However, it would be difficult to explain the lower velocity dispersions found in the region between the two galaxies. We cannot draw general conclusions from a single system that may even represent the exception instead of the rule but, if our interpretation is correct, this system may be one of the earliest cases of environmental effects known. In any case, our findings demonstrate that JWST may help long-standing problems in the local Universe to be solved. 

Finally, the NIRSpec IFS observations of GS\_4891 provide evidence of gas metallicities varying significantly within single systems at high redshifts. This fact is very relevant for studies based on multi-object spectroscopic observations that only probe regions of galaxies falling within the slits and not the whole systems, indicating that variations in metallicity within single systems might also contribute to the observed scatter between galaxies.



\subsubsection{Ionisation parameter}
\label{subsubsec:results_ionparameter}

Our dataset also allows the exploration of the ionisation parameter, $U$, which is the ratio of the ionising photon density to the particle density and which provides a measure of the degree of ionisation of the gas. This parameter can be estimated directly through O32, following Equation~7 in \citet{diazChemicalAbundancesIonizing2000}; however, since this indicator is used in the computation of the gas metallicity, the use of the previous equation leads by default to an anti-correlation between log$U$ and the gas metallicity. Instead, we employed Equation~5 from \citet{diazChemicalAbundancesIonizing2000}, which combines the gas metallicity, which has been well constrained above using several indicators, with the O2 ratio.
A map of log$U$ is shown in panel d) of Figure~\ref{fig:maps_derived_parameters}, whereas the integrated values for the three galaxies in the group and the two stellar clumps are included in Table~\ref{table:integrated_properties}.

\begin{figure}
\begin{center}
\includegraphics[width=0.45\textwidth]{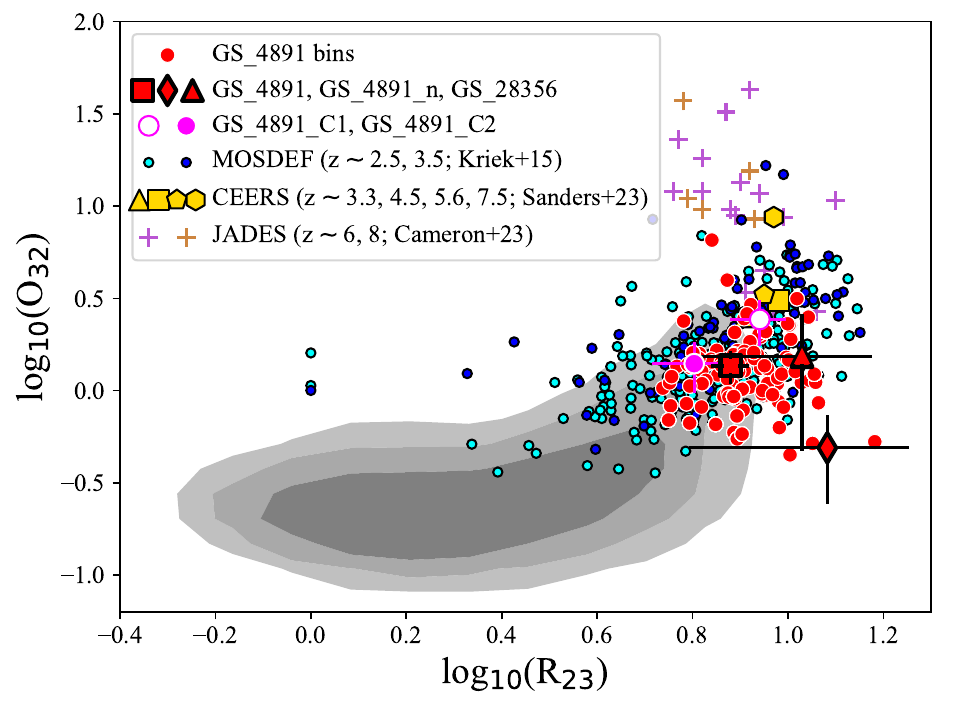}
\caption{R23$-$O32 diagram for the GS\_4891 system in comparison with other works. Spatially binned regions are shown as red dots, whereas integrated values are shown as red polygons and empty and filled magenta circles. As a reference, we have added the stacks of galaxies at $z\sim2.5, 3.5$ from \citet[][light and dark blue points, respectively]{kriekMOSFIREDEEPEVOLUTION2015}, the stacks in different redshift bins from \citet[][yellow symbols]{sandersExcitationIonizationProperties2023}, and the individual galaxies at $z\sim6, 8$ from \citet[][purple and brown pluses, respectively]{cameronJADESProbingInterstellar2023}. Shaded contours demarcate the regions in the diagram encompassed by 70, 80, and 90 percent of the local SDSS sample \citep{abazajianSeventhDataRelease2009}.}
\label{fig:R23_vs_O32}
\end{center}
\end{figure}

GS\_4891 presents the highest log$U$ values $>-2.6$ of the system in the south, in the region around the C1 clump, the centre, and in the northwest. In  GS$\_$4891$\_$n the values are generally lower, log$U<-2.6$, whereas GS$\_28356$ displays an intermediate value of log$U$=$-2.8$. Although log$U$ is found to anti-correlate with gas metallicity in local galaxies \citep[e.g.][]{perez-monteroDerivingModelbasedTeconsistent2014, sanchezImprintsGalaxyEvolution2015},
such behaviour is less clear at higher redshifts \citep{stromMeasuringPhysicalConditions2018, reddyImpactStarFormationRateSurface2023}. In GS\_4891 we observe anti-correlations between log$U$ and metallicity in clump C1, where log$U$ peaks and the metallicity is the lowest of the system. However, there are regions of high metallicity, such as the northwest of the galaxy, that also have a relatively high ionisation parameter. These results suggest that the physical conditions to hold an anti-correlation between log$U$ and metallicity might be present in individual regions of high-redshift galaxies, but not in the rest of the system. 

\begin{figure*}
\centering
\includegraphics[width=\textwidth]
{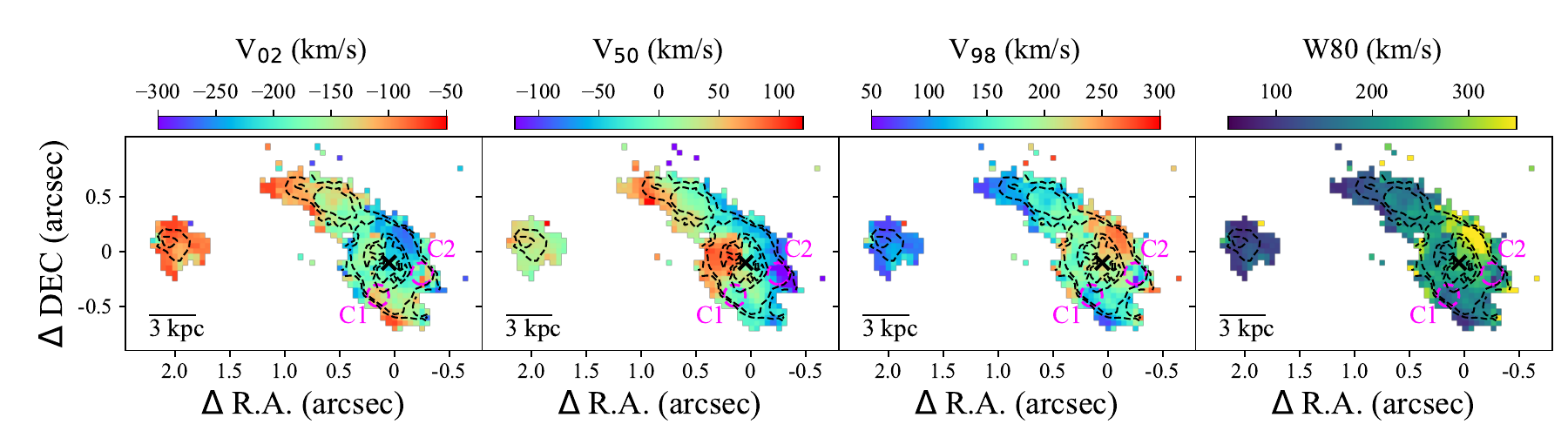}
\caption{Velocity maps of the ionised gas in the GS$\_$4891 system obtained from the fitting of the \OIIIb and \Ha lines. The three left panels show the velocities at the 2nd, 50th, and 98th percentiles of the velocity distributions. The right panel shows the W80 map. The black cross indicates the centroid of continuum
emission, whereas the dashed black contours correspond to the \Ha
emission}
\label{fig:kinematic_maps}
\end{figure*}

In Figure~\ref{fig:R23_vs_O32} we compare R23 and O32 in the different regions of our galaxies with those found in local galaxies \citep{abazajianSeventhDataRelease2009} at $z\sim 2$ \citep{sandersMOSDEFSurveyElectron2016} and up to $z>9.5$ \citep{cameronJADESProbingInterstellar2023,sandersExcitationIonizationProperties2023}. First of all, it is worth noticing the wide area of the diagram covered by the different regions in our galaxies, encompassing more than 0.2 and 0.5~dex in R$_{\rm23}$ and O$_{\rm32}$, respectively. Despite the large errors, GS\_4891\_n (red diamond) stands out from the rest of the sample due to its low O32 but high R23. In comparison with galaxies at $2<z<4$, regions in GS\_4891 are spread similarly in O32 but are more concentrated towards higher values of R23. Such high values of R23 are also shared by the high-redshift population, but they present $\geq0.2$~dex higher O32. Considering that the ionisation parameter correlates with O32 and the gas metallicity anti-correlates with R23, the results shown in Figure~\ref{fig:R23_vs_O32} 
indicate that the GS\_4891 galaxy group at $z\sim3.7$ is at an intermediate stage sharing roughly similar metallicities with the higher-redshift sample but lower ionisation, whereas galaxies at $z\leq4$ continue evolving towards lower ionisation but their gas metallicities increase.

\subsection{Ionised gas kinematics}
\label{subsec:results_gas_kinematics}
\subsubsection{Global kinematics}
\label{subsubsec:results_gas_kinematics}

We extracted the kinematic information of the ionised gas from the spectral modelling of the \OIIIb and \Ha lines performed in Section~\ref{subsec:analysis} tying together their velocities. These are the strongest lines in the spectra and allow us to constrain the extended kinematics throughout our system. Figure~\ref{fig:kinematic_maps} contains the velocity maps at the 2th, 50th, and 98th percentiles ($V_{02}$,$V_{50}$,$V_{98}$) and the velocity dispersion traced by the non-parametric estimate $W80$ (the line width that contains 80\% of the flux, obtained as the difference between the velocities at the 90th and 10th percentiles). First of all, the velocity maps confirm that the three galaxies lie at the same redshift, indicating that they belong to a common group. In the mid panel of Figure~\ref{fig:kinematic_maps}, the $V_{\rm50}$ map shows velocity gradients in GS$\_$4891 and GS$\_$4891$\_$n suggestive of rotating gas in both systems. As was already identified in Figure~\ref{fig:clumps_flux_maps}, clump C2 is moving at blue-shifted velocities of more than 100~km/s with respect to GS\_4891, whereas clump C1 displays similar kinematics to the rest of the galaxy. The maximum $V_{\rm50}$ differences are 175~km/s in GS\_4891 and 98~km/s in GS\_4891\_n. 
The $W80$ map shows a clear increase in the northwest region of GS\_4891, reaching values higher than $300$~km/s, indicative of higher turbulence, compared to the lower values observed in other regions of GS$\_$4891 and GS$\_$4891$\_$n. {This increase in $W80$ suggests that the ionised outflow detected in the nuclear parts of the galaxy potentially extends towards the northwest, a scenario that is explored in detail in the next section.} GS$\_$28356 displays a small variation in velocities (<50~km/s) and low $W80$ ($<150$~km/s). 

Motivated by the velocity gradients observed in GS$\_$4891, we attempted to model the gas kinematics with $\textsc{3D-BAROLO}$ \citep{diteodoro3DBAROLONew2015} following the method described in \citet{pernaPhysicsULIRGsMUSE2022}. From this analysis (presented in Appendix~\ref{a:3dbarolo}) we obtained a rotational velocity, $v_{\rm rot}$, of $\sim120$~km/s and a velocity dispersion of the gas, $\sigma_{\rm gas}$, of $\sim90$~km/s at a radius of 2.2~kpc. Using Equation 3 in \citet{pernaPhysicsULIRGsMUSE2022}, we first estimated the circular velocity, $v_{\rm c}$$\sim180$~km/s, assuming an asymmetric drift term $\eta$=3; then, using Equation 4 and assuming that the source of the gravitational potential is spherically distributed, we estimated that the dynamical mass within a radius of 2.2~kpc is $M_{\rm dyn}$$\sim$~$1.7\times$10$^{10}$~M$_{\odot}$. {We estimated a stellar mass within the area used for the kinematic modelling of $5.5\pm0.3\times10^9$~${\rm M}_{\odot}$, $\sim2$ times smaller than the dynamical mass, indicating that this galaxy is dominated by dark matter}.

\subsubsection{Detection of an ionised outflow}
\label{subsubsec:results_outflow}

\begin{figure}
\begin{center}
\includegraphics[width=0.49\textwidth]{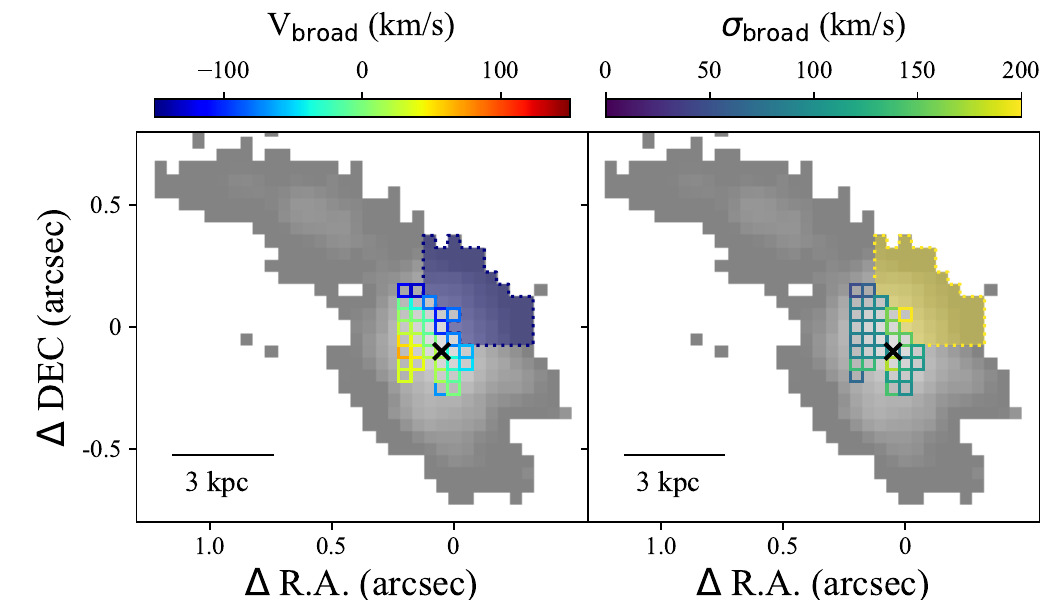}
\caption{Velocity and velocity dispersion maps of the broad kinematic component superimposed on the \Ha flux distribution in GS$\_$4891 (grey background). The black cross indicates the centroid of continuum emission.}
\label{fig:outflow_maps}
\end{center}
\end{figure}

Studies of galaxies at $z\geq1$ have revealed that galactic outflows are a relatively common phenomenon in massive ($\geq10^{10}{\rm M}_{\odot}$) galaxies \citep{pernaLBTARGOSAdaptive2018, forsterschreiberKMOS3DSurveyDemographics2019b, swinbankEnergeticsStarburstdrivenOutflows2019a,concasBeingKLEVERCosmic2022}, although at these masses outflows are more generally found in AGN systems \citep[e.g.][]{concasBeingKLEVERCosmic2022}. The presence of ionised outflows in the GS\_4891 system is explored through the modelling of the spectra using two kinematic components, as was described in Section~\ref{subsec:analysis}. This analysis reveals the presence of a second, broader kinematic component in several individual spaxels around the nuclear parts of GS$\_$4891. The maps of the mean velocity ($v_{\rm broad}$) and velocity dispersion ($\sigma_{\rm broad}$) of the broad component are shown in Figure~\ref{fig:outflow_maps}, superimposed on the \Ha flux distribution in GS$\_$4891. {Motivated by the observed increase in $W80$ towards the northwest of GS\_4891, we performed an additional kinematic modelling of the integrated spectrum extracted from the region with $W80>200$~km/s, highlighted with filled, dashed contours in the maps. In this integrated spectrum we also find evidence for a secondary, broad component in the \OIIIb and \Ha lines according to the criteria defined in Section~\ref{subsec:analysis}}. 

{The region where the outflow was detected in individual spaxels spans a total of $\sim4.4$~kpc$^2$, from which we estimated a circularised radius of the outflow of ~1.2~kpc. However, given that a broad component was detected in the integrated region to the northwest, we considered this radius to be a lower limit and adopted an alternative radius of the outflow, $r_{\rm out}=1.5$~kpc, which corresponds to the distance from the centre of the galaxy to half the width (approximately two spaxels) of the integrated region across the northwest direction. The velocity of the broad gas component, $v_{\rm broad}$, shows a gradient from the southeast ($\sim50$~km/s) to the northwest ($\sim-90$~km/s) of the galaxy centre, whereas the velocity dispersion, $\sigma_{\rm broad}$, is fairly constant throughout the extension of the outflow and only increases in the integrated region to the northwest.} The extension and kinematic properties indicate that the broad emission is likely associated with a blue-shifted, galactic outflow originating at the centre of GS$\_$4891. 
This outflow would have a maximum projected velocity, $v_{\rm out}$, of $\sim400$~km/s, estimated as the highest value of $\langle v_{\rm broad}\rangle + 2\sigma_{\rm broad}$ \citep[e.g.][]{fioreAGNWindScaling2017, carnianiJADESIncidenceRate2023, ublerGANIFSMassiveBlack2023a}, considering only the {regions} where a broad component is detected.

\begin{figure}
\begin{center}
\includegraphics[width=0.49\textwidth]{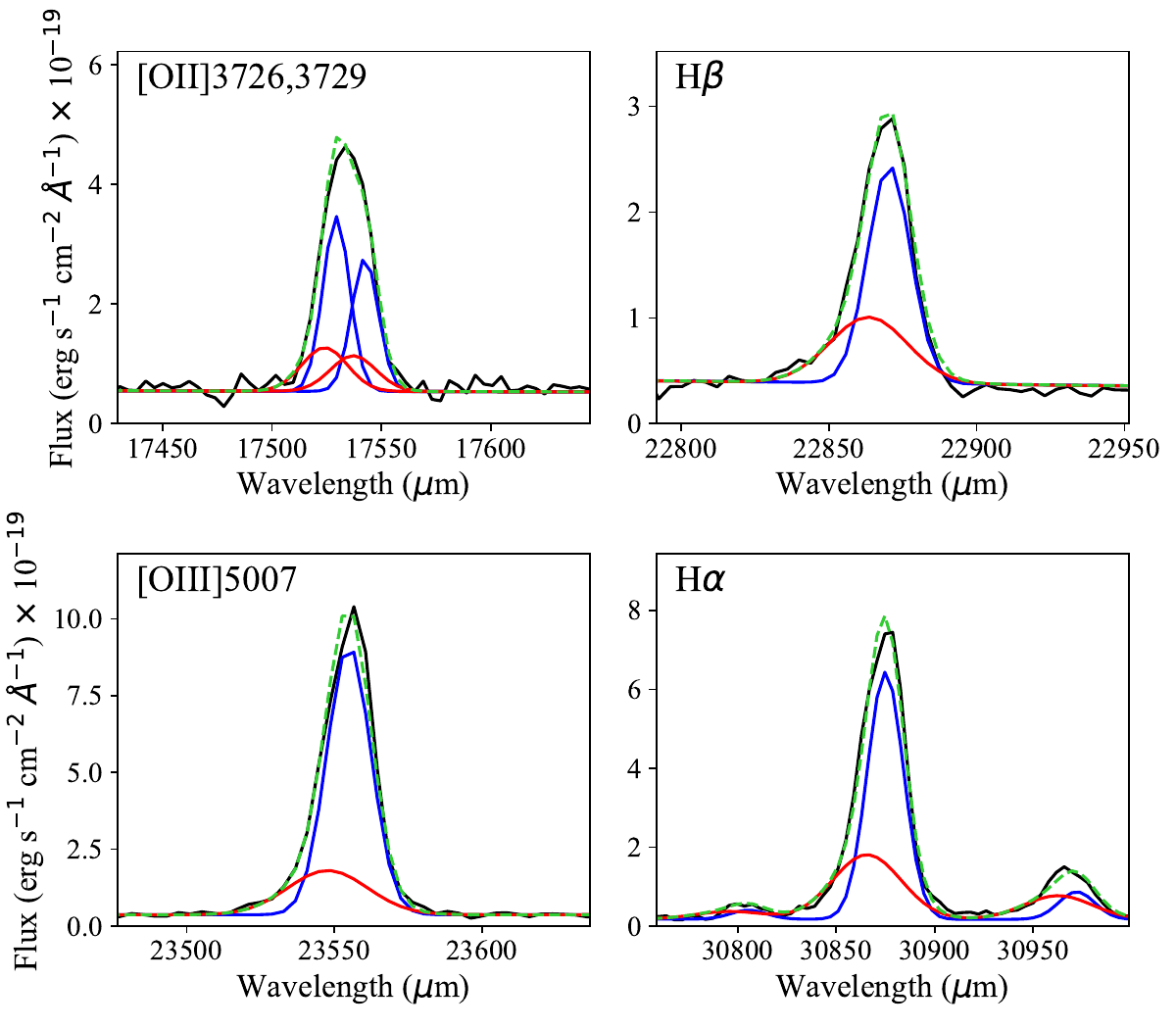}
\caption{Spectral modelling of the main emission lines in the integrated spectrum of the region hosting the ionised outflow. The blue and red lines correspond to the narrow and broad components, respectively, which summed together (green line) provide the best fit to the data (black line).}
\label{fig:outflow_fits}
\end{center}
\end{figure}

We explored the properties of the {outflowing gas} by integrating all the individual spaxels where a broad component is detected. {The region to the northwest was not included because the lower S/N in emission lines different to \OIIIb and \Ha reduces the quality of the fit. These line fluxes were used to estimate the nebular attenuation ($A_{\rm V, out}$) and the gas metallicity (12+log(O/H)$_{\rm out}$) in the outflow, whereas the associated $SFR$ and kinematic properties were estimated including the region to the northwest also}. We performed the two kinematic component modelling of this integrated spectrum of the outflow in the \OIIlines, \Hb, \OIIIb, \Ha and \NIIlines emission lines, which is shown in Figure~\ref{fig:outflow_fits}. This spectral modelling allowed us to constrain the broad component flux in the different lines that, together with the main parameters derived for the outflow, are reported in Table~\ref{table:kinematic_properties}. As can be inferred from the line fluxes, the contributions of the broad component to the total line flux in \OIIIb and \Ha are 23\% and 33\%, respectively, in agreement with the values found in ionised outflows of {massive galaxies ($>10^{11}$M$_{\odot}$) at $0.6 \leq z \leq 2.7$ \citep{forsterschreiberKMOS3DSurveyDemographics2019b, concasBeingKLEVERCosmic2022} and lower-mass ($<10^{10}$M$_{\odot}$) systems at $z$=$3-4$ \citep{guptaMOSELSurveyExtremely2023, llerenaIonizedGasKinematics2023}}.

In the BPT diagram shown in Figure~\ref{fig:BPT}, the fluxes in the broad component place the ionisation of the outflowing gas (blue circle) in a region of higher ionisation than local, star-forming galaxies but compatible with ionisation by star formation at the redshift of GS\_4891 (see  Section~\ref{subsubsec:results_BPT}). Under the assumption of star formation-driven ionisation of the outflowing gas, we can estimate its metallicity using the same prescriptions from \citet{curtiNewFullyEmpirical2017} as for the host galaxy (Section~\ref{subsubsec:results_metallicities}), in this case with the N2, R2, R3, R23, and O32 indicators. We obtain 12+log(O/H)~$\sim8.5_{-0.1}^{+0.1}$, which is $0.2$~dex higher than the one measured in the integrated spectrum of GS\_4891 (Secttion~\ref{table:integrated_properties}).

The properties of the outflow and its impact on the host galaxy can be further characterised with the estimation of the outflowing mass rate, $\dot{M}_{\rm out}$ (the rate at which the gas is being ejected by the outflow), and the energetics associated. The total mass of ionised gas associated with the outflow is estimated following Equation 1 from \citet{cresciBubblesOutflowsNovel2023}, which uses the \Ha luminosity of the broad component and the electron density of the outflow, $N_e$. In our data, the only set of emission lines that can be used to constrain the electron density in the outflow is \OIIlines, since the \SIIlines doublet presents low S/N to be modelled with two kinematic components. We note here that the NIRSpec spectral resolution barely allows us to resolve the \OIIlines and that constraining the fluxes of each kinematic component is very challenging. Therefore, instead of estimating the electron density from the broad component, we assume that the outflowing gas has the same electron density as all the gas in GS\_4891 ($n_{\rm e}$ = $776\pm307$~cm$^{-3}$) estimated in Section~\ref{subsubsec:results_ne}. 

{Thus, taking into account all the light within the region hosting the ionised outflow (individual spaxels and the integrated region to the northwest), we derive an outflowing gas mass, { $M_{\rm g}$~=~$\sim$~$1.3$~$\times$~$10^7$~M$_{\odot}$}. Considering  the maximum velocity of the outflowing gas ($v_{\rm out}$) and the size of the outflow ($r_{\rm out}$), we estimate an outflowing mass rate, $\dot{M}_{\rm out}\sim4$~ M$_{\odot}$/yr. Based on the \Ha flux and nebular extinction ($1.0\pm0.3$~mag, derived as in Section~\ref{subsubsec:extinction}) of the total (narrow + broad) line fluxes, we estimate an associated SFR, following \citet{kennicuttStarFormationGalaxies1998}, of $57.6\pm7.4$~M$_{\odot}$/yr. This SFR is higher than the one reported for GS\_4891 in Table~\ref{table:integrated_properties} ($37.5\pm0.6$~M$_{\odot}$/yr), because the area spanned by the outflow is larger than the one defined to extract the integrated spectrum from GS\_4891 (Figure~\ref{fig:flux_maps}) and now we are also including the \Ha flux in the broad component. The ratio between the outflowing mass rate and the total SFR in the host region, known as the mass-loading factor ($\eta_{\rm out}$=$\dot{M}_{out}$/$SFR$), is equal to $\sim0.04$}. This low $\eta$ value indicates that the ionised outflow is not having a significant impact on the star formation activity in the host galaxy, although we have to bear in mind that we are only tracing the ionised phase of the gas and more outflow mass could be in colder and hotter gas phases \citep[e.g.][]{belliMassiveMultiphaseGas2023}. {Our measured $\eta$ is in agreement with those reported in massive ($>10^{11}$M$_{\odot}$) star-forming galaxies at $0.6 \leq z \leq 2.7$ \citep[$\eta<0.2$; ][]{forsterschreiberKMOS3DSurveyDemographics2019b, concasBeingKLEVERCosmic2022} but lower than in less massive ($<10^{10}$M$_{\odot}$) systems at $z$~=~$3-4$ \citep[$0.05<\eta<3.26$;][]{guptaMOSELSurveyExtremely2023, llerenaIonizedGasKinematics2023} and at $z=3-9$ \citep[$0.5<\eta<3.6$;][]{carnianiJADESIncidenceRate2023}}. {Finally, the corresponding kinetic power of the ionised outflow is $\dot{E}_{\rm kin}$=$1.8\times10^{41}$ erg s$^{-1}$ and the momentum rate, $\dot{P}_{\rm kin}$=$9.2\times10^{33}$ dyne. As was mentioned in Section~\ref{subsubsec:results_ne}, the ionised outflow might also contribute to an increase in the electron density of the gas in the galaxy, which could explain the high electron densities measured in GS\_4891}. 

Finally, we explored the possibility that the outflowing gas might escape from the host galaxy by estimating the escape velocity of the system using the dynamical mass estimated in Section~\ref{subsubsec:results_gas_kinematics}. Following \citet{arribasIonizedGasOutflows2014}, we computed the escape velocity at {$r_{\rm out}$~=~1.5~kpc} for an isothermal sphere truncated at 2.2~kpc (the radius within which the dynamical mass is computed), {obtaining $v_{\rm esc}$~$\sim$~240~km/s}. Given that $v_{\rm out}$ can reach values up to 
$\sim400$~km/s, at least part of the outflowing gas could escape the galaxy or be re-distributed throughout the galaxy. This result would imply that outflows might also be contributing to the regulation of metals at $z$~=3.7. Moreover, given that the outflow is oriented towards the northwest of GS\_4891, in the case in which the gas again falls on the galaxy (in the form of a galactic fountain; \citealt{fraternaliAccretionGasNearby2008}), it could contribute to explaining the observed positive metallicity gradient from the nucleus of GS\_4891 to the northwest region (see panel c) in Fig.~\ref{fig:maps_derived_parameters}). Thus, the inside-out distribution of metals via outflows would provide an alternative scenario to explain positive metallicity gradients in high-redshift galaxies, in addition to inflows of pristine gas directly into the central regions \citep{cresciGasAccretionOrigin2010, curtiKLEVERSurveySpatially2020}. Nevertheless, as was discussed in Section~\ref{subsubsec:results_metallicities}, in the case of GS\_4891 the higher metallicities at the northwest could also be a consequence of the interaction with GS\_4891\_n, which leads to the mixing of high-metallicity gas. 

\begin{table}
\caption{{Properties of the spatially resolved outflow in GS\_4891.}}
\begin{tabular}{lcc}
\hline 
\hline
\rule{0pt}{2.5ex}  
Line Fluxes
($\times$ 10$^{\rm -19}$ erg s$^{-1}$ cm$^{-2}$)\\
\hline
 & Narrow & Broad \\
\hline
\rule{0pt}{2.5ex}  
\OIIa  & $45\pm5$ & $19\pm5$ \\
\OIIb  & $34\pm6$ & $16\pm8$ \\
\Hb    & $38\pm4$ & $21\pm4$ \\
\OIIIb & $163\pm8$ & $49\pm9$\\
\Ha    & $144\pm11$ & $71\pm11$\\ 
\NIIb  & $16\pm3$ & $26\pm4$\\
\hline
\hline
\end{tabular} 
\centering
\rule{0pt}{2.5ex}  
\centerline{Derived outflow parameters}
\begin{tabular}{lc}
\hline
\rule{0pt}{2.5ex}  
\rule{0pt}{2.5ex}  
$r_{\rm out}$ (kpc) & $1.5$ \\
$V_{\rm out}$ (km/s) & $-90$, $50$ \\
$\sigma_{\rm out}$ (km/s) & $68$, $198$  \\
$A_{\rm V, out}$ (mag)  & $1.0\pm0.3$ \\
12+log(O/H)$_{\rm out}$ & $8.5_{-0.1}^{+0.1}$ \\
$\dot{M}_{\rm out}$ (M$_{\odot}$yr$^{-1}$) & 4  \\ 
$\eta_{\rm out}$ & $0.04$ \\
\hline
\hline
\end{tabular} 
\label{table:kinematic_properties}
\end{table}

\section{Summary and conclusions}
\label{sec:conclusions}
In this work we present the JWST/NIRSpec IFS data of a highly star-forming galaxy group around the massive GS\_4891 galaxy at $z\sim3.7$, observed as part of the GA-NIFS program. We used the low-resolution data (R100) to estimate stellar masses, whereas we used the high-resolution (R2700) observations that include a wealth of emission lines, from \OIIlines to \SIIlines, to study with an unprecedented spatial and spectral resolution the ISM properties and ionised gas kinematics of GS\_4891 and its close environment within a region of 20x20~kpc$^2$. The wide range of emission lines in our spectra have allowed us to study the spatially resolved distribution of the nebular extinction through the Balmer decrement, the excitation mechanisms through standard BPT diagnostic diagrams, the gas metallicity using a total of five indicators, and the ionisation parameter. We have also studied the kinematics of the galaxy group, from which we have estimated the dynamical mass of GS\_4891 and detected the presence of an ionised outflow in this galaxy. Our main findings are: 

\begin{itemize}  
\item GS\_4891 is part of a small group of galaxies together with one galaxy to the north, GS\_4891\_n, which could be interacting with GS\_4891, and another one to the east, GS\_28356, at a projected distance of less than 15~kpc. Additionally, GS\_4891 displays evidence of internal substructure in the form of two star-forming clumps in the southeast and west that could be indicative of minor mergers and/or gas accretion. 

\item The nebular extinction estimates in GS\_4891, computed using the \Hd, \Hg, \Hb, and \Ha hydrogen Balmer emission lines, are in agreement with local attenuation curves \citep{cardelliRelationshipInfraredOptical1989, calzettiDustContentOpacity2000}, suggesting that the effects of dust
along the line of sight at $z\sim3.7$ are similar to those in the local Universe. 

\item The excitation mechanisms throughout GS\_4891 and GS\_4891\_n are consistent with star formation, with no signs of AGN activity. These systems occupy regions in the BPT and R23$-$O32 diagrams in between $z\sim2$ and $z>4$ galaxies, consistent with an increase in the ionisation parameter and a decrease in gas metallicity with redshift. 

\item The gas metallicity in GS\_4891\_n is 0.2dex higher than at the centre of GS\_4891 despite being four times less massive. There is a drop in metallicity to the south of GS\_4891, close to clump C1, which we associate with a minor merger with a lower metallicity companion or with the accretion of relatively metal-poor gas. GS\_4891 presents a positive metallicity gradient ($0.15$dex) towards the northwest, resembling those observed at slightly lower redshifts \citep{curtiKLEVERSurveySpatially2020, cresciGasAccretionOrigin2010} and {recently also at higher redshifts \citep{arribasGANIFSCoreExtremely2023}, which are likely associated with inflows of cold gas in the central regions}. The $0.2$~dex difference in metallicity between GS\_4891 and GS\_4891\_n {could be} a consequence of inflows of pristine gas being more favoured in the more massive galaxy \citep[`starvation'][]{larsonEvolutionDiskGalaxies1980a, pengStrangulationPrimaryMechanism2015}, potentially leading to a differential build-up of the mass-metallicity relation between satellites and centrals. Assuming star formation is truncated earlier in satellites, these observations could help to explain the relative enrichment of satellite galaxies with respect to central galaxies of the same mass at low redshifts \citep{pasqualiGasphaseMetallicityCentral2012a, pengDependenceGalaxyMassmetallicity2014}. Preferential accretion of pristine gas by the central galaxy would also explain the positive metallicity gradient observed in GS\_4891. Alternative explanations to this positive gradient include the mixing of metal-rich gas from GS\_4891\_n and the distribution of metal-rich gas by the ionised outflow detected in GS\_4891. 

\item GS\_4891 hosts a metal-rich, ionised outflow around the central regions that extends out {to $r_{\rm out}$=$1.5$~kpc}. The ionised outflow reaches maximum velocities $r_{\rm out}\sim400$~km/s, yielding an outflowing mass rate, $\dot{M}_{\rm out}\sim4$, which is significantly smaller than the SFR in the host region, $57.6\pm7.4$~M$_{\odot}$/yr. The corresponding low mass-loading factor, $\eta\sim0.04$, implies that the outflow has little effect on the SFRs of the host. 
  
\end{itemize}

The results obtained in this work indicate that the system around GS\_4891 might be in a pre-merger stage, based on the large differences in metallicities and the lack of a high gas disturbance between GS\_4891 and GS\_4891\_n that would have been indicative of an ongoing merger. This work showcases the potential of NIRSpec IFS observations to study the internal structure and close environment of high-redshift systems, which is the main goal of the GA-NIFS survey. 


\begin{acknowledgements}
We would like to acknowledge the whole JWST mission and the instrument science teams for a successful launch and commissioning of the observatory. This work is based on observations made with the NASA/ESA/CSA \textit{James Webb} Space Telescope. The data were obtained from the Mikulski Archive for Space Telescopes at the Space Telescope Science Institute, which is operated by the Association of Universities for Research in Astronomy, Inc., under NASA contract NAS 5-03127 for JWST. These observations are associated with program \#1216, as part of the NIRSpec Galaxy Assembly IFS GTO program. 

BRP, MP and SA acknowledge grant PID2021-127718NB-I00 funded by the Spanish Ministry of Science and Innovation/State Agency of Research (MICIN/AEI/ 10.13039/501100011033). MP also acknowledges the Programa Atracci\'on de Talento de la Comunidad de Madrid via grant 2018-T2/TIC-11715. IL acknowledges support from PID2022-140483NB-C22 funded by AEI 10.13039/501100011033 and BDC 20221289 funded by MCIN by the Recovery, Transformation and Resilience Plan from the Spanish State, and by NextGenerationEU from the European Union through the Recovery and Resilience Facility. PGP-G acknowledges support from grants PGC2018-093499-B-I00 and PID2022-139567NB-I00 funded by Spanish Ministerio de Ciencia e Innovaci\'on
MCIN/AEI/10.13039/501100011033, FEDER, UE. GC, GV and MP acknowledge the support of the INAF Large Grant 2022 "The metal circle: a new sharp view of the baryon cycle up to Cosmic Dawn with the latest generation IFU facilities". GV and SC acknowledge support from the European Union (ERC, WINGS, 101040227). AJB, GCJ and JC acknowledge funding from the "FirstGalaxies" Advanced Grant from the European Research Council (ERC) under the European Union’s Horizon 2020 research and innovation programme (Grant agreement No. 789056).
H{\"U} gratefully acknowledges support by the Isaac Newton Trust and by the Kavli Foundation through a Newton-Kavli Junior Fellowship. FDE, JS and RM acknowledge support by the Science and Technology Facilities Council (STFC), from the ERC Advanced Grant 695671 "QUENCH". FDE and JS also acknowledge support by the UKRI Frontier Research grant RISEandFALL. RM also acknowledges funding from a research professorship from the Royal Society. This research made use of Astropy, a community-developed core Python package for Astronomy \citep{astropy_collaboration_astropy_2018}. This research has made use of the NASA/IPAC Extragalactic Database (NED) which is operated by the Jet Propulsion Laboratory, California Institute of Technology, under contract with the National Aeronautics and Space Administration. This work has made use of the Rainbow Cosmological Surveys Database, which is operated by the Centro de Astrobiología (CAB), CSIC-INTA, partnered with the University of California Observatories at Santa Cruz (UCO/Lick,UCSC).
\end{acknowledgements}

%
%
\bibliographystyle{aa}
\bibliography{4891_edited}


\appendix

\section{3D-Barolo fit}\label{a:3dbarolo}
\begin{figure}
\centering
\includegraphics[width=0.49\textwidth]
{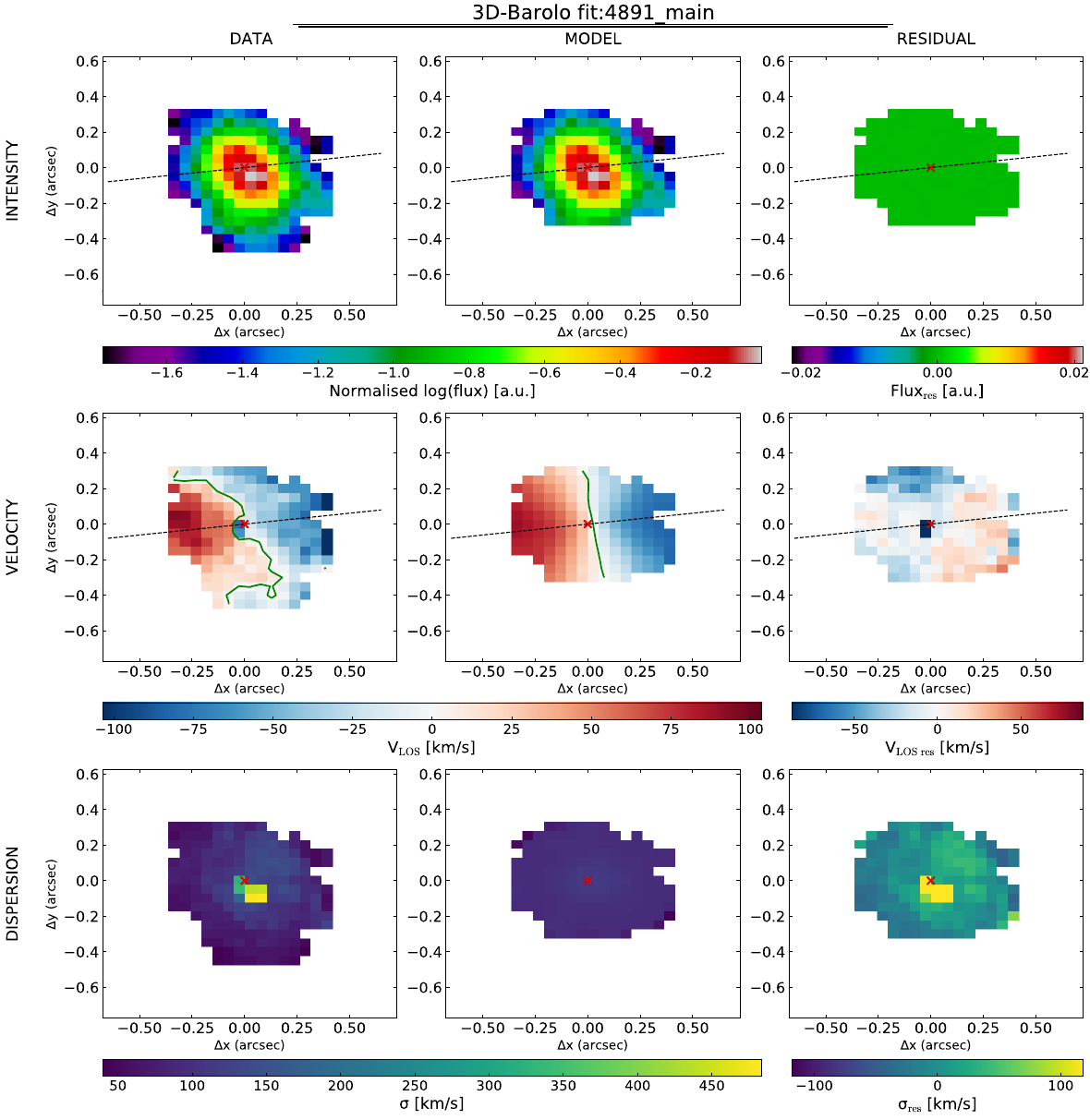}
\caption{Disc kinematic best-fit of the total flux, velocity, and velocity dispersion of GS\_4891 (first to third rows). The analysis was performed with \textsc{3D-Barolo} \citep{diteodoro3DBAROLONew2015} on the kinematics maps obtained from a spaxel-by-spaxel emission-line fitting (see Section~\ref{subsec:results_gas_kinematics} and Figure~\ref{fig:kinematic_maps}). The black and green lines identify the major axis and the zero-velocity curve, respectively.}
\label{fig:barolo}
\end{figure}

The best-fit model of the ionised gas kinematics in GS\_4891 performed with \textsc{3D-Barolo} is shown in Figure~\ref{fig:barolo}. The velocity maps can be well reproduced with a rotating disc model. 

\end{document}